\documentclass[pra,onecolumn,superscriptaddress,aps,5pt]{revtex4}
\usepackage{amsfonts}
\usepackage{amssymb,latexsym,amsmath}
\usepackage{graphicx}
\usepackage{dcolumn}
\usepackage{times}
\usepackage{subfigure}
\usepackage[toc,page,title,titletoc,header]{appendix}
\usepackage[usenames]{color}
\usepackage{ulem}

\newcommand{\wo}{\omega_0}
\newcommand{\wm}{\omega_\mathrm{m}}
\newcommand{\sech}{\mathrm{sech}}

\begin{document}

\title{Beating dark-dark solitons in Bose-Einstein condensates}

\author{D.~Yan}
\affiliation{Department of Mathematics and Statistics,
University of Massachusetts,
Amherst, Massachusetts 01003-4515, USA}
\author{J.J.~Chang}
\affiliation{Washington State University,
Department of Physics \& Astronomy,
Pullman, Washington 99164, USA}
\author{C.~Hamner}
\affiliation{Washington State University,
Department of Physics \& Astronomy,
Pullman, Washington 99164, USA}
\author{M. Hoefer}
\affiliation{North Carolina State University,
Department of Mathematics and Statistics,
Raleigh, North Carolina 27695, USA}
\author{P.G.~Kevrekidis}
\affiliation{Department of Mathematics and Statistics, University of Massachusetts,
Amherst, Massachusetts 01003-4515, USA}
\author{P.~Engels}
\affiliation{Washington State University,
Department of Physics \& Astronomy,
Pullman, Washington 99164, USA}
\author{V.~Achilleos}
\affiliation{Department of Physics, University of Athens, Panepistimiopolis,
Zografos, Athens 157 84, Greece}
\author{D.J.~Frantzeskakis}
\affiliation{Department of Physics, University of Athens, Panepistimiopolis,
Zografos, Athens 157 84, Greece}
\author{J. Cuevas}
\affiliation{Nonlinear Physics Group.
Escuela Polit\'{e}cnica Superior.
Departamento de F\'{\i}sica Aplicada I.
Universidad de Sevilla.
C/ Virgen de \'{A}frica, 7. 41011-Sevilla (Spain)}

\begin{abstract}


Motivated by recent experimental results, we
%consider
study beating dark-dark
solitons as a prototypical coherent structure that emerges in
two-component Bose-Einstein condensates. We showcase their connection
to dark-bright solitons via SO(2) rotation, and infer from it both
their intrinsic beating frequency and their frequency of oscillation
inside a parabolic trap. We identify them as exact periodic orbits in
the Manakov limit of equal inter- and intra-species nonlinearity
strengths with and without the trap and showcase the persistence of
such states upon weak deviations from this limit. We also consider
large deviations from
%this limit,
the Manakov limit illustrating that this breathing
state may be broken apart into dark-antidark soliton states. Finally, we
consider the dynamics and interactions of two beating dark-dark solitons in
the absence and in the presence of the trap, inferring their typically
repulsive interaction.

\end{abstract}

\maketitle

\section{Introduction}

One of the principal themes of study in the emerging field of atomic
Bose-Einstein condensates (BECs) is the examination of the coherent
structures that arise in them~\cite{emergent,revnonlin,rab,djf}.
When such explorations started over a decade
ago~\cite{han1,nist,dutton,bpa,han2}, they were considerably
hindered by either geometric or thermal effects, which were
detrimental towards the lifetime of dark solitons and vortices
%the solitary wave and vortex states
that can be formed in
%of interest within the realm of
repulsive BECs. Yet, the newer generations of experiments
have enabled considerable strides towards the observation of dynamics
and interactions of such nonlinear waveforms~\cite{hamburg,hambcol,technion,kip,kip2,engels}.

\begin{figure}[tbp]
\centering
\includegraphics[width=8.5cm]{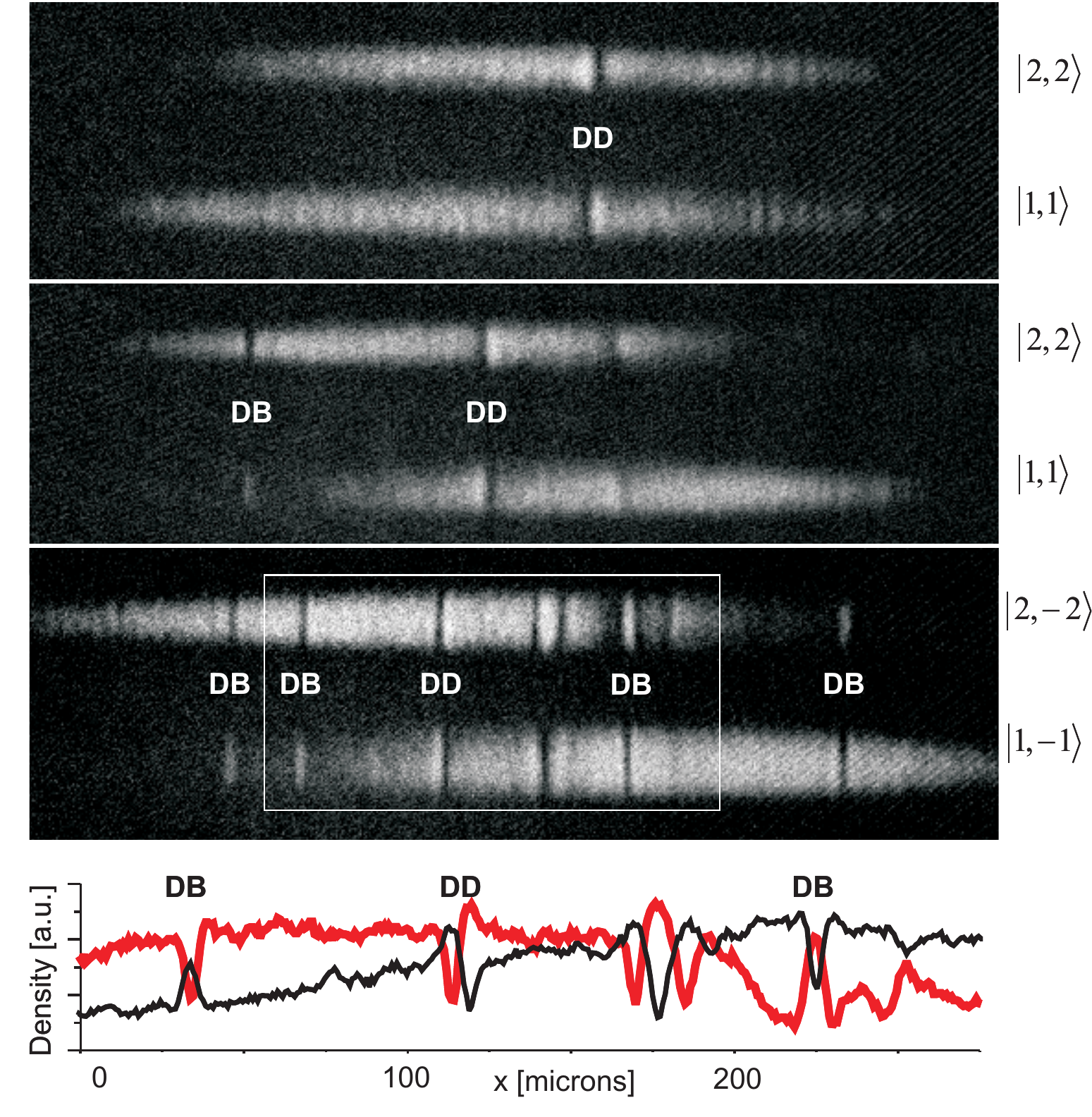}
\caption{(Color online) Prototypical experimental images of
dark-bright and dark-dark solitons in a two-component BEC. The two
components are vertically offset for separate imaging. All dynamics
occur with vertically overlapped components before the imaging
procedure. Clear examples of dark-bright and dark-dark solitons are
marked as DB and DD respectively. In the fourth panel, the red
(thick) line shows a radially integrated cross section of the upper
component in the boxed region of the third panel, while the black
(thin) line shows the cross section of the lower component. The
$|F,m_{F}\rangle$ hyperfine states used for these images are given
to the right of each component.}
 \label{experiment}
\end{figure}

In addition to the above context of single-component BECs, soliton
and vortex states may also arise in multi-component condensates,
such as the two-component pseudo-spinor BECs, or the three and
higher component spinor BECs~\cite{emergent,revnonlin,djf}. One of
the prototypical examples of a soliton state in these settings
%such a waveform
is the so-called dark-bright (DB) soliton~\cite{BA,DDB}.
Experimental images of DB solitons in a two-component BEC are
presented in Fig.~\ref{experiment}. The BEC in this figure is
comprised of two different hyperfine states of $^{87}$Rb, and the
solitons are generated by subjecting the BEC to inter-component
counterflow; details of this technique are described
in Refs.~\cite{engels1,engels3}. In each panel, the atom clouds of the two
components are vertically offset for imaging only, while all the
dynamics leading to the soliton formation occurs in overlapped
clouds. Clear examples of dark-bright solitons are marked as ``DB''
in the figure and they consist of a dark soliton in one component that
is coupled to a bright soliton in the second component.
%They
These structures can be
thought of as ``symbiotic'' (or even parasitic) states because their
bright component cannot be supported alone in the case of repulsive
interactions \cite{rab}; in fact, the bright soliton is only
sustained because of the presence of its dark-counterpart, which
operates as an external trapping potential.
Although
%the observation of
dark-bright solitons (and even a prototypical interaction thereof)
%was first reported
were first observed some time ago in the context of nonlinear optics~\cite{seg1,seg2},
%and this led to some analytical investigations of their profile~\cite{ablowitz},
their observation in recent atomic BECs experiments~\cite{hamburg}
%but it was not until they were observed in atomic BECs~\cite{hamburg} that
triggered a sizeable burst of research activity
%was
centered around them.
Topics of study included (but were not limited to) multi-DB soliton
solutions from the viewpoint of integrable systems~\cite{rajendran},
numerical study of DB soliton interactions~\cite{berloff}, discrete DB
solitons~\cite{azucena},
%as well as
experimental realizations of DB soliton trains~\cite{engels1},
DB soliton oscillations and interactions~\cite{engels2,engels4},
as well as interaction of DB solitons with localized impurities \cite{vas}.

Recently, a ``cousin'' of these DB solitons, namely the dark-dark (DD)
soliton --- which involves two dark solitons but with potentially
a breathing oscillation between their densities
was also experimentally
%identified
observed~\cite{engels3}. Pertinent examples are
%the features
marked as ``DD''
in Fig.~\ref{experiment}. These solitons show interesting dynamics
in which they periodically change their form, from the one shown in
the first panel to the one shown in the second panel, and back (note
the order of the hump/notch features in each of the DD's component;
see also Fig.~\ref{fig:perturb2} below). Such ``beating dark-dark
solitons'' are expected to emerge in the integrable two-component
(so-called Manakov) limit of the relevant mean-field theoretic
models~\cite{parkshin} and were, in fact, earlier observed in
numerical experiments involving the dragging of defects through the
binary condensates~\cite{hadi}.

The current experimental advances, such as the ones leading to the
soliton images of Fig.~\ref{experiment}, motivate the present
theoretical study, in which we revisit DD soliton states at the
integrable Manakov limit and extract information from their
connection to the DB solitons (section II). These results are
corroborated by the identification of such single DD soliton states, as
genuine periodic orbits of the Manakov case (with and without a
trap) and the
%examination
study of their stability, internal modes and
associated near-equilibrium dynamics (section III). In addition, we
examine the dynamics of individual such solitons, upon departure
from the integrable limit (section IV).
%The stability and dynamics of beating DD solitons are studied as
%periodic orbits arising in the Manakov limit (section IV), while
%Finally,
Experimentally it has also become possible to generate several
solitons, and even solitons of different types, in a single BEC
-- see,
%. As an example,
e.g., the third panel of Fig.~\ref{experiment} which demonstrates the coexistence
%coexisting
of dark-bright and dark-dark solitons.
%coexisting in a single BEC.
%While
Although the experimentally exploited counterflow between the two
components is beyond the scope of our current analysis,
these experimental findings
%possibilities nevertheless
motivate our investigation of the
interactions between two dark-dark solitons (section V). Finally,
conclusions of our study, as well as
%Section V concludes our study and offers
a number of interesting perspectives for future work, are also
presented (section VI).

\section{Dark-bright and dark-dark solitons:
%single dark-bright and dark-dark soliton states in two-component Bose-Einstein condensates:
theoretical background}

We consider a two-component elongated (along the $x$-direction) repulsive BEC, composed of two different hyperfine states of the same alkali isotope. In the case of a highly anisotropic trap (i.e., if the longitudinal and transverse trapping frequencies are such that $\omega_x \ll \omega_{\perp}$), this system can be described by two coupled Gross-Pitaevskii equations (GPEs) of the form \cite{emergent}:
\begin{eqnarray}
i\hbar \partial_t \psi_j =
%\nonumber \\
\left( -\frac{\hbar^2}{2m} \partial_{x}^2 \psi_j +V(x) -\mu_j + \sum_{k=1}^2 g_{jk} |\psi_k|^2\right) \psi_j.
\label{model}
\end{eqnarray}
Here, $\psi_j(x,t)$ ($j=1,2$) denote the mean-field wave functions of the two components (normalized to the numbers of atoms $N_j = \int_{-\infty}^{+\infty} |\psi_j|^2 dx$), $m$ is the atomic mass, and $\mu_j$ are the chemical potentials; furthermore,
$g_{jk}=2\hbar\omega_{\perp} a_{jk}$ are the effective one-dimensional (1D) coupling constants, $a_{jk}$ denote the three $s$-wave scattering lengths (note that $a_{12}=a_{21}$) which account for collisions between atoms belonging to the same ($a_{jj}$) or different ($a_{jk}, j \ne k$) species, while $V(x)=(1/2)m\omega_x^2 x^2$ is the external trapping potential.

Let us now assume that the two-component BEC under consideration consists of two different hyperfine states of $^{87}$Rb, such as the states $|1,-1\rangle$ and $|2,1\rangle$ used in the experiment of Ref.~\cite{mertes}, or the states $|1,-1\rangle$ and $|2,-2\rangle$ used in the experiments of Refs.~\cite{engels1,engels2,engels3}. In the first case the scattering lengths take the values $a_{11}=100.4a_0$, $a_{12}=97.66a_0$ and $a_{22}=95.00a_0$, while in the second case the respective values are $a_{11}=100.4a_0$, $a_{12}=98.98a_0$ and $a_{22}=98.98a_0$ (where $a_0$ is the Bohr radius). In either case, it is clear that the scattering lengths take approximately the same values, say $a_{ij} \approx a$. This way, measuring the densities $|\psi_j|^2$, length, time and energy in units of $2a$, $a_{\perp} = \sqrt{\hbar/\omega_{\perp}}$, $\omega_{\perp}^{-1}$ and $\hbar\omega_{\perp}$, respectively, we may cast Eqs.~(\ref{model}) into the following dimensionless form,
\begin{eqnarray}
i \partial_t u_1  &=& -\frac{1}{2} \partial_{x}^2 u_1  + V(x)u_1
%\nonumber \\
+(|u_1|^2 + |u_2|^2 -\mu) u_1,
\label{deq1}
\\
i\partial_t u_2  &=& -\frac{1}{2} \partial_{x}^2u_2 +V(x)u_2
%\nonumber \\
+ (|u_1|^2 + |u_2|^2- \mu) u_2,
\label{deq2}
\end{eqnarray}
where we have also assumed that the chemical potentials characterizing each component are equal. Note that the potential in Eqs.~(\ref{deq1})-(\ref{deq2}) is now given by $V(x)=(1/2) \Omega^2 x^2$, where $\Omega = \omega_x/\omega_{\perp}$ is a natural small parameter of the system.

%The two-component vector nonlinear Schr{\"o}dinger (NLS) equation modeling the
%(1+1)d dynamics where the states of interest herein prototypically arise
%is:
%%
%\begin{eqnarray}
%i\partial_t u_1 &=& -\frac{1}{2}\partial_{xx}u_1+V(x)u_1+\left(g_{11}|u_1|^2+g_{12}|u_2|^2-\mu\right)u_1
%\label{u_1}
%\\
%i\partial_t u_2 &=& -\frac{1}{2}\partial_{xx}u_2+V(x)u_2+\left(g_{12}|u_1|^2+g_{22}|u_2|^2-\mu\right)u_2
%\label{u_2}
%\end{eqnarray}
%
%where we have also assumed that the chemical potentials characterizing each component are
%equal. Note that the potential in Eqs.~(\ref{u_1})-(\ref{u_2}) is given by
%$V(x)=\frac{1}{2}\Omega^2x^2$. Also

The above system of Eqs.~(\ref{deq1})-(\ref{deq2}) is
invariant under SU(2) rotations~\cite{parkshin}. In particular, let us first recall that a general matrix
element of SU(2) takes the form
\[ U=\left( \begin{array}{ccc}
\alpha & -\beta^{*} \\
\beta & \alpha^{*} \end{array} \right),\]
where $\alpha$ and $\beta$ are complex constants such that
$|\alpha|^2+|\beta|^2=1$. Then, it can be shown that if
$(u_1,u_2)^{T}$ are solutions of Eqs.~(\ref{deq1})-(\ref{deq2}), then,
\[ \left( \begin{array}{ccc}
u_1'\\
u_2' \end{array} \right)
\equiv U \left( \begin{array}{ccc}
u_1 \\
u_2 \end{array} \right)
= \left( \begin{array}{ccc}
\alpha u_1-\beta^{*} u_2\\
\beta u_1+\alpha^{*} u_2 \end{array} \right),
\]
are also solutions of Eqs.~(\ref{deq1})-(\ref{deq2}). This suggests
that we may start from the exact dark-bright (DB) soliton solution
(which exists in the absence of the potential) and, derive the beating
dark-dark (DD) soliton solution. More specifically, in the absence of
the external potential ($V(x)=0$), and for the boundary conditions
$|u_1|^2\rightarrow\mu$ and $|u_2|^2\rightarrow 0$ as $|x|\rightarrow
\infty$, Eqs.~(\ref{deq1})-(\ref{deq2}) possess an exact analytical
%one
single DB soliton solution of the following form:
\begin{eqnarray}
u_1(x,t) &=& \sqrt{\mu} \{\cos\phi\tanh \xi+i \sin\phi\},
\label{dark_part}
\\
u_2(x,t) &=& \eta\, \sech\, \xi \exp\{ikx+i\theta(t)\},
\label{bright_part}
\end{eqnarray}
where $\xi=D(x-x_0(t))$, $\phi$ is the dark soliton's phase angle,
$\cos\phi$ and $\eta$ represent the amplitude of the dark and bright
solitons, and $D$ and $x_0(t)$ are associated with the inverse width
and the center position of the DB soliton. Furthermore, $k=D\tan\phi$
and $\theta(t)$ are the (constant) wavenumber and phase of the bright
soliton, respectively. The above parameters of the DB soliton are
connected through the following equations:
\begin{eqnarray}
D^2 &=& \mu\cos^2\phi-\eta^2,
\label{D}
\\
\dot{x}_0 &=& k= D\tan\phi,
\label{x_0}
\\
\dot{\theta} &=& \frac{1}{2}\left(D^2-k^2\right),
\label{theta}
\end{eqnarray}
with $\dot{x}_0$ and $\dot{\theta}$ denoting the DB soliton
velocity and angular frequency, respectively (overdots denote time derivatives).
Thus, the DB soliton \eqref{dark_part}, \eqref{bright_part} is characterized by three
free parameters (seven parameters $\mu$, $\phi$, $\eta$, $k$, $D$,
$\dot{x}_0$, $\dot{\theta}$ and four constraints \eqref{D}-\eqref{theta}).
Notice that the amplitude $\eta$ of the
bright soliton, the chemical potential $\mu$ of the dark soliton, as
well as the (inverse) width parameter $D$ of the DB soliton are
connected to the number of atoms $N_B$ of the bright soliton by means
of the following equation:
\begin{eqnarray}
N_B \equiv \int_{\mathbb{R}}|u_2|^2\mathrm{d}x=\frac{2\sqrt{\mu}\eta^2}{D}.
\label{N_B}
\end{eqnarray}
According to the above arguments, one may start from the DB soliton
and construct SU(2) rotated solutions, in the following form:
\begin{eqnarray}
  u_1(x,t) &=&
  \alpha\sqrt{\mu}\{\cos\phi\tanh\xi+i\sin\phi\}-\beta^{*}\eta \,
  \sech \, \xi \exp\{ikx+i\theta(t)\},
  \label{new_u1}
  \\
  u_2(x,t) &=&
  \beta\sqrt{\mu}\{\cos\phi\tanh\xi+i\sin\phi\}+\alpha^{*}\eta\,
  \sech\, \xi \exp\{ikx+i\theta(t)\}.
\label{new_u2}
\end{eqnarray}
With the additional four parameters $\alpha$, $\beta \in
  \mathbb{C}$ and the constraint $|\alpha|^2 + |\beta|^2 = 1$, the
  solution \eqref{new_u1}-\eqref{new_u2} is characterized by six free
  parameters.  Introducing a new parameter $c$, the velocity of the
  background fluid,
  %via a Galilean boost,
  another solution can be
  constructed from Eqs.~\eqref{new_u1}-\eqref{new_u2} via a Galilean boost:
  $\exp[i(cx-c^2t/2)] u_{1,2}(x-ct,t)$. Thus, in the most general
  case, this DD soliton solution is characterized by seven free
  parameters. One natural set of parameters can be found from the
  far-field, $|x| \to \infty$ behavior consisting of two densities, an
  overall fluid velocity, and four phases.
% It is important to mention that the parameters $\mu$, $D$, $\phi$,
% $\eta$, $k$, $\theta$ and $\dot{x}_0$ of the above DD soliton are
% still connected via Eqs.~(\ref{D})-({\ref{theta}}). Thus, the
% solution (\ref{new_u1})-(\ref{new_u2}) is characterized by 7 free
% parameters, since there exist, in total, 11 parameters and 4
% equations connecting them [Eqs.~(\ref{D})-({\ref{theta}}) and the
% constraint $|\alpha|^2+|\beta|^2=1$].  Furthermore, we should
% mention that still another additional parameter may exist by
% considering a Galilean boost of the solution
% (\ref{new_u1})-(\ref{new_u2}).  Indeed, introducing a new parameter
% $c$, which is defined as the velocity of the background fluid,
% another solution that can be constructed from
% Eqs.~(\ref{new_u1})-(\ref{new_u2}) is:
% $u_{1,2}(x-ct,t)\exp[icx-i(c^2/2)t]$; thus, in the most general
% case, the DD soliton solution is characterized by 8 free parameters.

Due to Galilean invariance and phase invariance, $u_j'(x,t) =
  e^{i \varphi_j} u_j(x,t)$, we
  will assume,
  %can,
  without loss of generality,
  %assume % To
% reduce the number of independent variables
% %appearing in Eqs.~(\ref{new_u1})-(\ref{new_u2}),
% below we will consider, for simplicity,
that the background is at rest ($c=0$) and
%we
focus, more
specifically, on the case of the SO(2)
rotated DB soliton. In this case, the corresponding orthogonal matrix
is given by:
\begin{equation}
U=\left( \begin{array}{ccc}
\cos(\chi) & -\sin(\chi) \\
\sin(\chi) & \cos(\chi) \end{array} \right),
\label{so2}
\end{equation}
where $\chi$ is an arbitrary angle. This way, the relevant SO(2)
rotated soliton solution takes the form:
\begin{eqnarray}
u_1(x,t) &=&
\cos(\chi)\sqrt{\mu}\{\cos\phi\tanh(D(x-x_0(t)))+i\sin\phi\}-\sin(\chi)\eta
\sech(D(x-x_0(t))) \exp\{ikx+i\theta(t)\},
\label{new_u1_sim}
\\
u_2(x,t) &=&
\sin(\chi)\sqrt{\mu}\{\cos\phi\tanh(D(x-x_0(t)))+i\sin\phi\}+\cos(\chi)\eta
\sech(D(x-x_0(t))) \exp\{ikx+i\theta(t)\},
\label{new_u2_sim}
\end{eqnarray}
The solution (\ref{new_u1_sim})-(\ref{new_u2_sim}) is a DD
  soliton solution characterized by 4 free parameters. The
asymptotics of this solution are $|u_1|^2\rightarrow \mu \cos^2(\chi)$
and $|u_2|^2\rightarrow \mu \sin^2(\chi)$ as
$|x|\rightarrow\infty$. The densities of the above dark solitons read:
\begin{eqnarray}
n_1 \equiv |u_1|^2 &=& \mu \cos^2(\chi) -(\mu \cos^2(\chi) \cos^2 \phi - \eta^2 \sin^2(\chi)){\rm sech}^2 \xi
\nonumber \\
&-& \sqrt{\mu} \eta \sin(2\chi) \left\{\sin\phi \sin[kx+\theta(t)] + \cos\phi \cos[kx+\theta(t)]\tanh \xi \right\}
{\rm sech}\xi,
\label{den1} \\
n_2 \equiv |u_2|^2 &=& \mu \sin^2(\chi) -(\mu \sin^2(\chi) \cos^2 \phi - \eta^2 \cos^2(\chi)){\rm sech}^2 \xi
\nonumber \\
&+& \sqrt{\mu} \eta \sin(2\chi) \left\{\sin\phi \sin[kx+\theta(t)] + \cos\phi \cos[kx+\theta(t)]\tanh \xi \right\}
{\rm sech}\xi,
\label{den2}
\end{eqnarray}
while the total density $n_{\rm tot}$ of the DD soliton is given by:
\begin{equation}
n_{\rm tot} = n_1+n_2 = \mu - D^2 {\rm sech}^2\xi.
\label{totden}
\end{equation}
Notice that the total density of the DD soliton is time-independent
and has the form of a dark soliton density of depth $D^2$ on top of a
background density $\mu$.
%Notice that
%This
The above density is, in fact,
identical to the density of the DB soliton; this is due to the fact
that under SO(2) rotation the total density, as well as all other
conserved quantities of the system, remain unchanged. This
%fact
will be particularly important when considering the motion of the DD
soliton in a trap --- see below.

On the other hand, one may consider the individual dark soliton densities, $n_1$ and $n_2$, across the trajectory of the DD soliton, i.e., for $\xi=0$: in such a case, $x=x_0(t)=kt$ and the densities read:
\begin{eqnarray}
n_1(\xi=0) &=& \mu \cos^2(\chi) \sin^2 \phi + \eta^2 \sin^2(\chi)
\nonumber \\
&-& \sqrt{\mu} \eta \sin(2\chi) \sin\phi \sin\left[\frac{1}{2}(k^2+ D^2)t \right],
\label{den10} \\
n_2(\xi=0) &=& \mu \sin^2(\chi) \sin^2 \phi + \eta^2 \cos^2(\chi)
\nonumber \\
&+& \sqrt{\mu} \eta \sin(2\chi) \sin\phi \sin\left[\frac{1}{2}(k^2+ D^2)t \right].
\label{den20}
\end{eqnarray}
It is clear that $n_{1,2}(\xi=0)$ are periodic functions of time; the relevant angular frequency
(which constitutes the internal beating frequency
of the DD soliton) is given by:
\begin{eqnarray}
\omega_0 = \frac{1}{2}(k^2+D^2)= \frac{1}{2}(\mu-\eta^2\sec^2 \phi),
\label{om}
\end{eqnarray}
where we have also used Eq.~(\ref{D}). The frequency $\omega_0$ is
bounded by two limiting values. First, in the case $\eta \rightarrow
0$, the beating DD soliton becomes a regular DD soliton, characterized
by a width $D=\sqrt{\mu}\cos\phi$ and a velocity $k=\sqrt{\mu} \sin
\phi$; in this case, $\omega_0 \rightarrow (1/2)\mu$. Second, in the
limiting case $D\rightarrow 0$, the beating DD soliton is reduced to a
plane wave; in this case, $\omega_0 \rightarrow (1/2)k^2$. In other
words, the intrinsic oscillation frequency take values in the range:
\begin{eqnarray}
\frac{1}{2}k^2 <\omega_0 < \frac{1}{2} \mu.
\label{limom}
\end{eqnarray}

\section{Dark-Dark solitons as periodic orbits in the Manakov model}
%:Numerical  Existence, Linear Stability and Near-Equilibrium Dynamics}

\begin{figure}
\begin{tabular}{cc}
 \includegraphics[width=6cm]{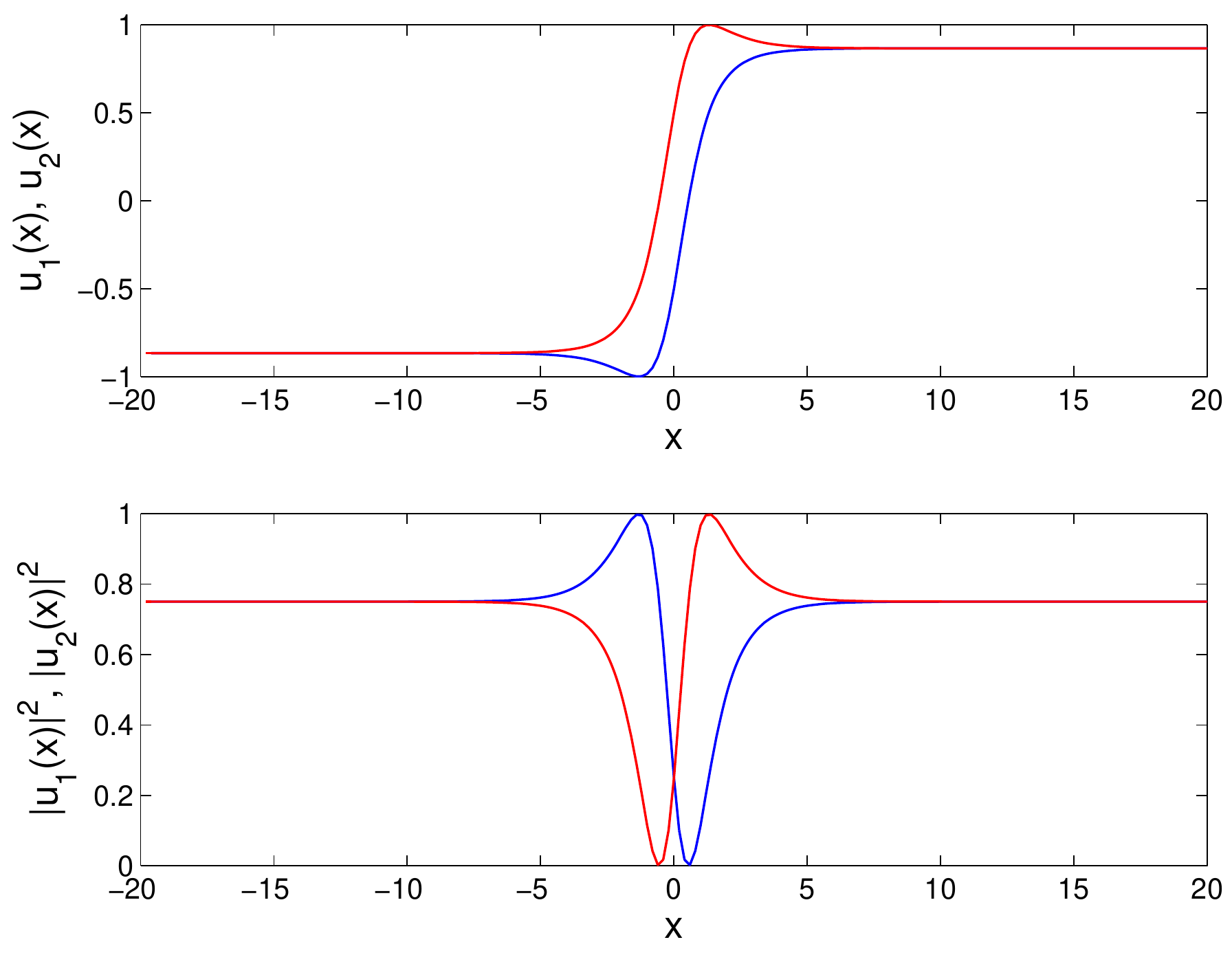} &
 \includegraphics[width=6cm]{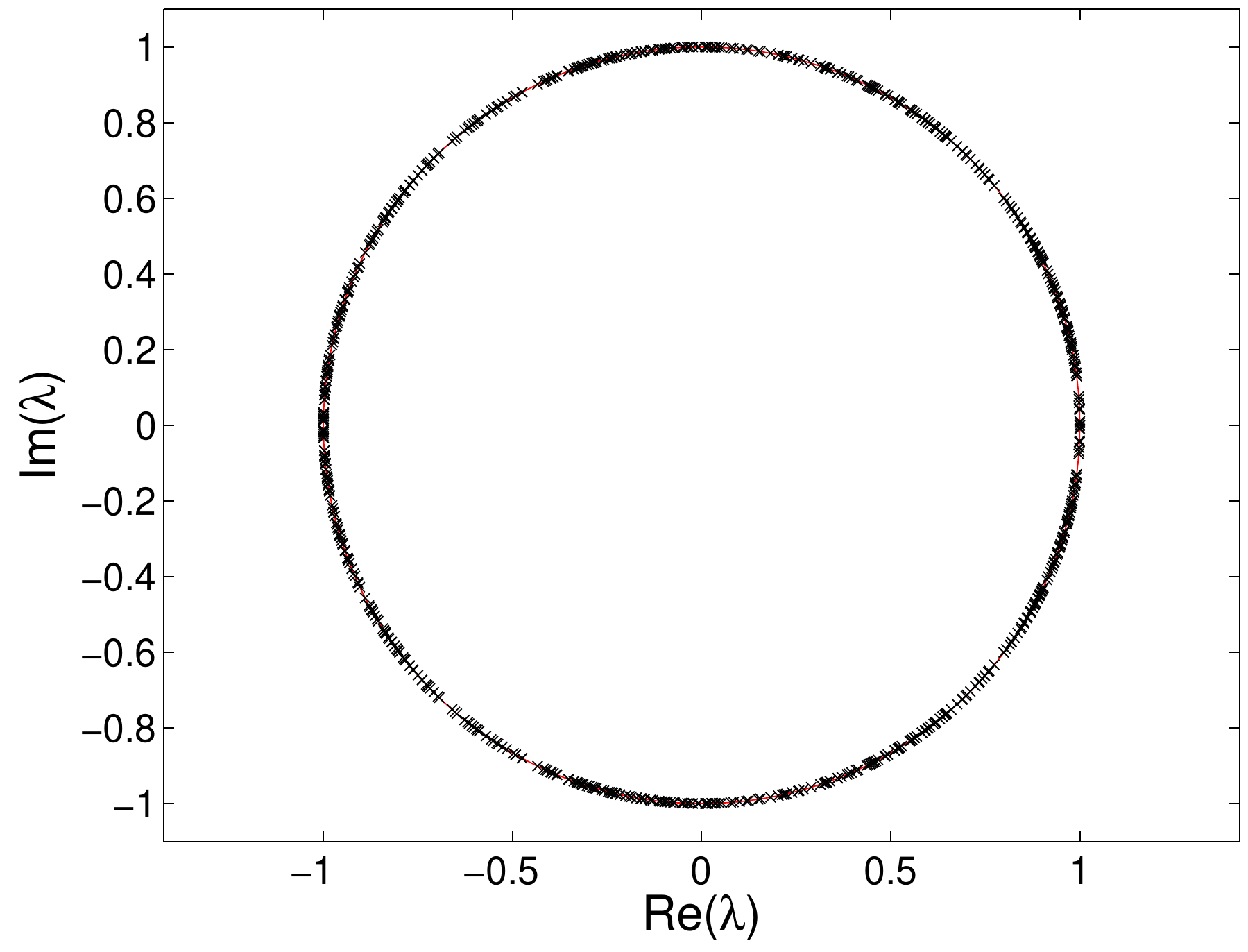} \\
\end{tabular}
\caption{Left panel: Profiles and densities of the beating dark-dark soliton
solution with $\Omega=0$, $\wo=0.5$ and $\mu=1.5$ at $t=0$. Right panel: Floquet multipliers spectrum for the dark-dark soliton displayed in the left panel.}
\label{fig:orbit_notrap}
\end{figure}

\begin{figure}
\begin{tabular}{cc}
 \includegraphics[width=6cm]{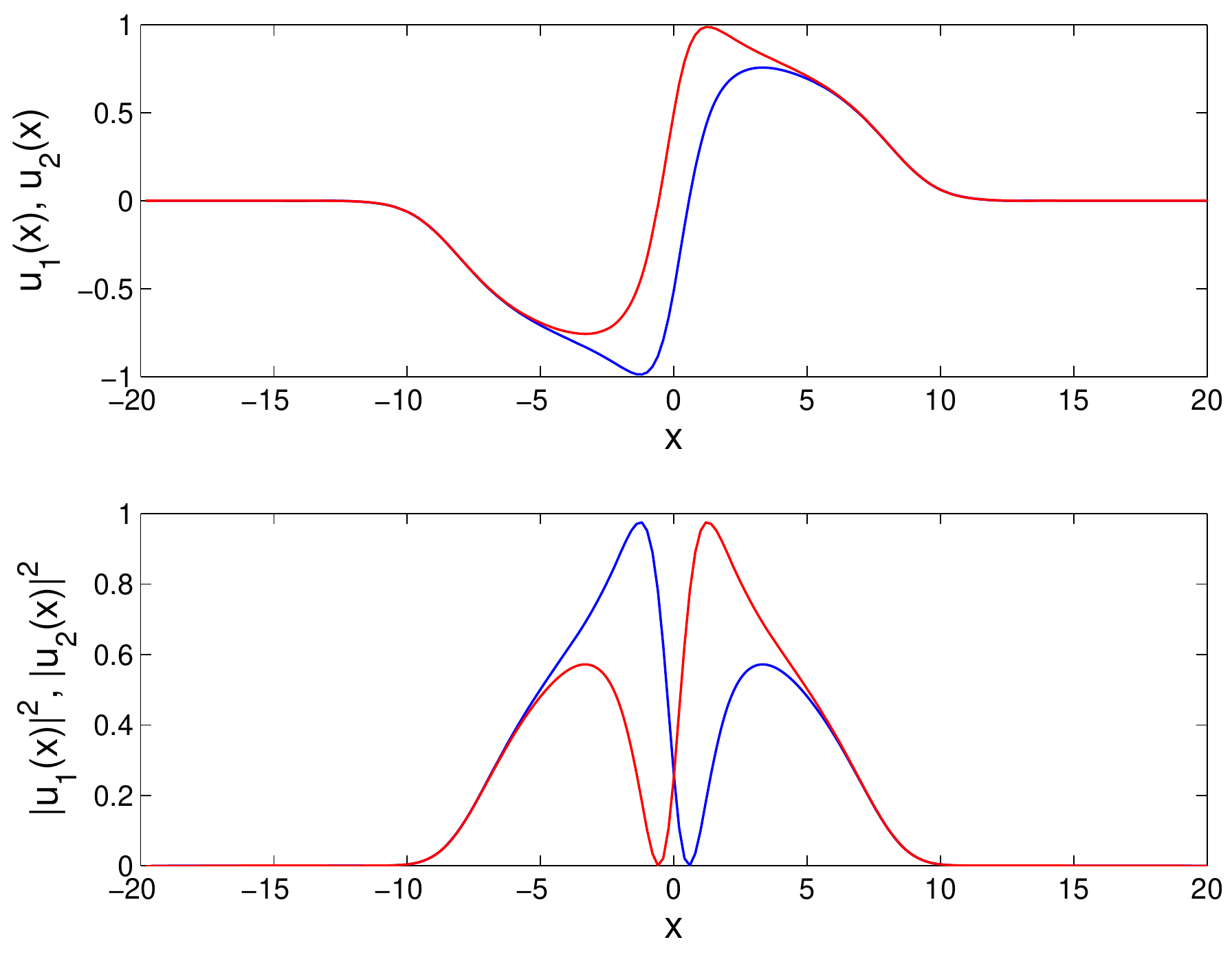} &
 \includegraphics[width=6cm]{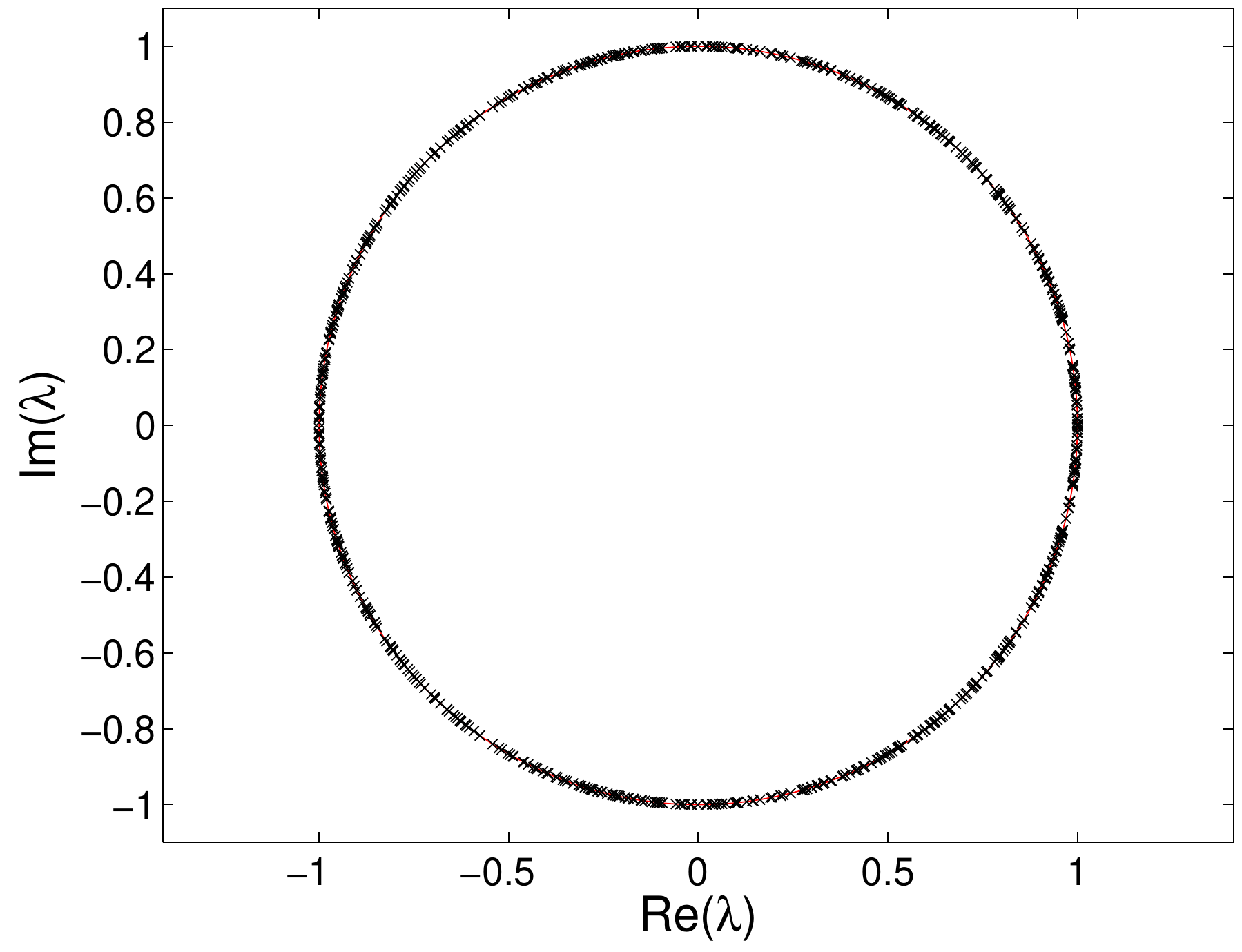} \\
\end{tabular}
\caption{Same as Fig. \ref{fig:orbit_notrap} but
%taking
in the trapped case with $\Omega=0.2$.}
\label{fig:orbit_trap}
\end{figure}

In this section, we analyze the existence, stability and dynamics of single beating DD
%dark-dark
solitons
%with
in a trap of the form $V(x)=\frac{1}{2}\Omega^2x^2$,
considering them as periodic orbits.
%Let us consider
%Now, we are about to study
%the single beating DD
%one dark-dark
%soliton solution within a harmonic trap of the form
%i.e., for
%$V(x)=\frac{1}{2}\Omega^2x^2$.
In the presence of the trap, the
dynamics of the center of mass $x_0(t)$ of the beating DD soliton is
still described by the dynamics of the original (unrotated) DB soliton
center $x_0$. This is due to the fact that the GPEs (\ref{deq1})-(\ref{deq2}) are invariant under SO(2) rotations even in the presence of $V(x)$, and so are all conserved quantities of the system, such as the total energy. Since the derivation of the equation of motion for the DB soliton center $x_0$ in Ref.~\cite{BA} was relying on the change of energy (due to the presence of the trap), it is clear that the evolution of the beating DD soliton center follows the same dynamics: it performs
a harmonic oscillation in the trap according to the equation $\ddot{x}_0+\omega_{osc}^2 x_0=0$,
where the oscillation frequency $\omega_{osc}$ is given by \cite{BA}:
\begin{eqnarray}
\omega_{osc}^2 &=&
\Omega^2\left(\frac{1}{2}-\frac{r}{8\sqrt{1+(\frac{r}{4})^2}}\right)
\label{omega_osc}
\end{eqnarray}
where $r=\frac{N_B}{\sqrt{\mu}}$ is a measure of the ratio of the number
of atoms in the bright and dark soliton component.  In order to compute
the soliton profile and determine its stability,
%To this end,
we consider the solution of Eqs.~(\ref{deq1})-(\ref{deq2}), with
$g_{11}=g_{22}=g_{12}=1$, as a Fourier series expansion of period
$\wo$ \cite{note}, namely,
%\footnote{Note that we cannot do this analysis for systems with a ratio between scattering lengths different to 1, as in that case, there is always an oscillation of the center of mass (alike to that of a particle in a well) with a frequency non-commensurable to the beating frequency. Consequently, there are no periodic orbits.}
%
\begin{equation}
    u_1(x,t)=\sum_{k=-\infty}^\infty z_k(x)\mathrm{e}^{ik\wo t},
    \qquad u_2(x,t)=\sum_{k=-\infty}^\infty y_k(x)\mathrm{e}^{ik\wo t},
\end{equation}
with $\{z_k\},\{y_k\}\in\mathbb{R}$. Then, the dynamical equations
%turn into
are reduced to a set of coupled equations:

\begin{eqnarray}
[\mu-k\wo-V(x)]z_k+\frac{1}{2}\partial^2_x z_k &=& \sum_p\sum_q (z_p z_q^*+y_p y_q^*) z_{k-p+q}
\label{Four1}
\\ [0pt]
[\mu-k\wo-V(x)]y_k+\frac{1}{2}\partial^2_x y_k &=& \sum_p\sum_q (z_p z_q^*+y_p y_q^*) y_{k-p+q}
\label{Four2}
\end{eqnarray}
where we have used the notation $z_k\equiv z_k(x)$, $y_k\equiv y_k(x)$.
%In the case of absence of trap,
If the trap is absent, it is straightforward to see that
\begin{eqnarray}
    z_0(x)&=&\sqrt{\frac{\mu}{2}}\tanh(\sqrt{2\omega_0}x)=y_0(x),
    \\
    z_1(x)&=&-\sqrt{\frac{\mu}{2}-\omega_0}\sech(\sqrt{2\omega_0}x)=-y_1(x),
\\  z_j(x)&=&y_j(x)= 0, \qquad |j| >1 \quad or \quad j=-1,
\end{eqnarray}
%
%which
is actually the solution (\ref{new_u1_sim})-(\ref{new_u2_sim})
for $\chi=\pi/4$, $\phi=k=0$, and $\omega_0=D^2/2$. In order to
%calculate the beating
%dark-dark
numerically find a DD soliton solution in the system with the trap, the previous solution
(with the dark component $\{z_k\}$ multiplied by the Thomas-Fermi cloud)
is introduced as a seed for a fixed-point method in the system of
%equations
Eqs.~(\ref{Four1})-(\ref{Four2}). Throughout this section, we have considered ---for convenience--- a trap strength
%of the trap
$\Omega=0.2$ in order to consume less time in
%simplify
the numerical calculations. Figures~\ref{fig:orbit_notrap} and \ref{fig:orbit_trap} show the periodic orbit for $t=0$
%and, respectively,
without and with a trap potential, respectively. It is worth remarking that solutions in the trap exist for $\mu>2\wo$, as predicted in the end of section II.
%the previous sections.

Once a periodic solution is found, its (linear) orbital stability can be analyzed by means of Floquet analysis. To this end, the time evolution of a small perturbation ${\xi_1(x,t),\xi_2(x,t)}$ to a periodic solution $\{u_{1,0}(x,t),u_{2,0}(x,t)\}$ must be traced. These perturbations are introduced in the dynamical equations (\ref{deq1})-(\ref{deq2}) as:
\begin{equation}
    u_1(x,t)=\left[u_{1,0}(x,t) + \delta \xi_1(x,t)\right], \qquad
    u_2(x,t)=\left[u_{2,0}(x,t) + \delta \xi_2(x,t)\right],
\end{equation}
and the resulting equation
%fulfilled
at order $O(\delta)$ reads:
%by the perturbations is:
%
\begin{eqnarray}
i \partial_t
\xi_1=[-\frac{1}{2}\partial^2_x+2|u_{1,0}|^2+|u_{2,0}|^2-\mu+V(x)]\xi_1+u_{1,0}^2\xi_1^*
+u_{2,0}^*u_{1,0}\xi_2+u_{2,0}u_{1,0}\xi_2^*,
\label{perturb1}
\\
i \partial_t
\xi_2=[-\frac{1}{2}\partial^2_x+|u_{1,0}|^2+2|u_{2,0}|^2-\mu+V(x)]\xi_2+u_{2,0}^2\xi_2^*
+u_{1,0}^*u_{2,0}\xi_1+u_{1,0}u_{2,0}\xi_1^*.
\label{perturb2}
\end{eqnarray}

Then, in the framework of Floquet analysis, the stability properties
of periodic orbits
%as can be determined by means of a Floquet analysis, . It
are resolved by diagonalizing the monodromy matrix $\mathcal{M}$,
which is defined as:
\begin{equation}
\left[\begin{array}{c}
  \mathrm{Re}(\xi_1(x, T)) \\ \mathrm{Im}(\xi_1(x, T)) \\ \mathrm{Re}(\xi_2(x, T)) \\ \mathrm{Im}(\xi_2(x, T)) \\
  \end{array}\right]
  =\mathcal{M}
  \left[\begin{array}{c}
  \mathrm{Re}(\xi_1(x, 0)) \\ \mathrm{Im}(\xi_1(x, 0)) \\ \mathrm{Re}(\xi_2(x, 0)) \\ \mathrm{Im}(\xi_2(x, 0)) \\
  \end{array}\right] .
\end{equation}
with $T=2\pi/\wo$. As the system is symplectic and Hamiltonian, the
linear stability of the solutions requires that the monodromy
eigenvalues, $\lambda$ (also called Floquet multipliers) must lie on
the unit circle (see, e.g., \cite{hadijesus,Melvin} for details). The
Floquet multipliers can also be written as $\lambda=\exp(i\Theta)$,
with $\Theta$ being denoted as the Floquet exponent. An internal mode
of the soliton corresponds to a spatially localized solution of
  Eqs.~\eqref{perturb1}-~\eqref{perturb2},
with its oscillation frequency related to the Floquet exponents
as
%
%\begin{equation}
    $\wm=\Theta \wo/(2\pi)$.
%\end{equation}
%
%\edit{[Question: What is the purpose of introducing $\wm$ since it is
%  not referred to again?  What is it the frequency of?]}
Figures~\ref{fig:orbit_notrap} and \ref{fig:orbit_trap} show a typical
Floquet multiplier spectra, indicating stability of the periodic
orbits. All the analyzed solutions (i.e. with $\Omega=0$ and
$\Omega=0.2$) are stable.

\begin{figure}
\begin{tabular}{cc}
 \includegraphics[width=6cm]{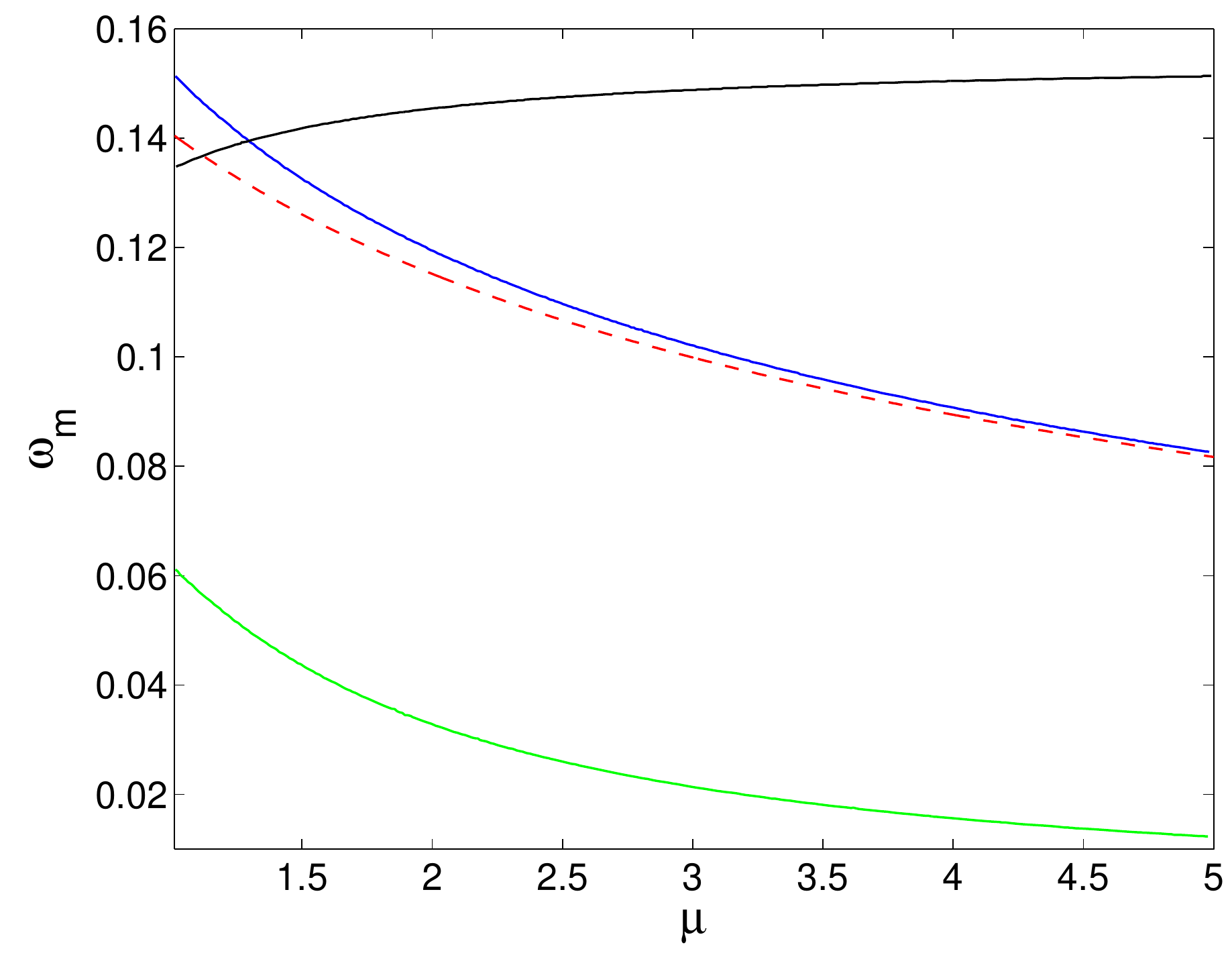} &
 \includegraphics[width=6cm]{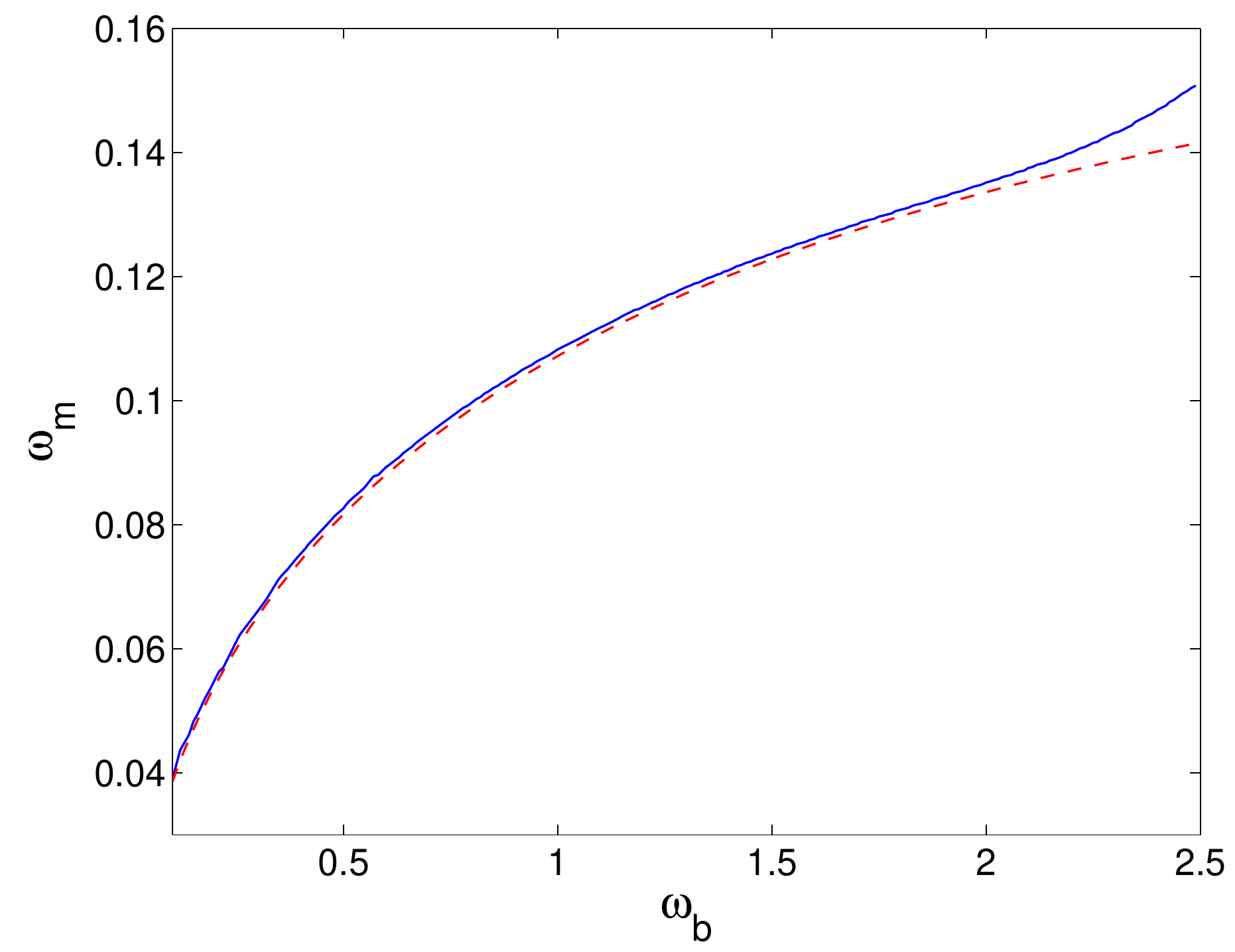} \\
\end{tabular}
\caption{(Left panel) Dependence of the eigenfrequencies of some internal modes (solid lines) with respect to $\mu$ with fixed $\wo=0.5$ (The black, blue, green and red lines are the upper solid, middle solid, lower solid and the dashed respectively). The  right panel
shows that dependence with respect to $\wo$ with fixed $\mu=5$ for the mode in blue in the left panel. In both panels, the red dashed line corresponds to the oscillation frequency (\ref{omega_osc}) in a trap with $\Omega=0.2$.}
\label{fig:modes1}
\end{figure}

\begin{figure}
\begin{tabular}{ccc}
 \includegraphics[width=4cm]{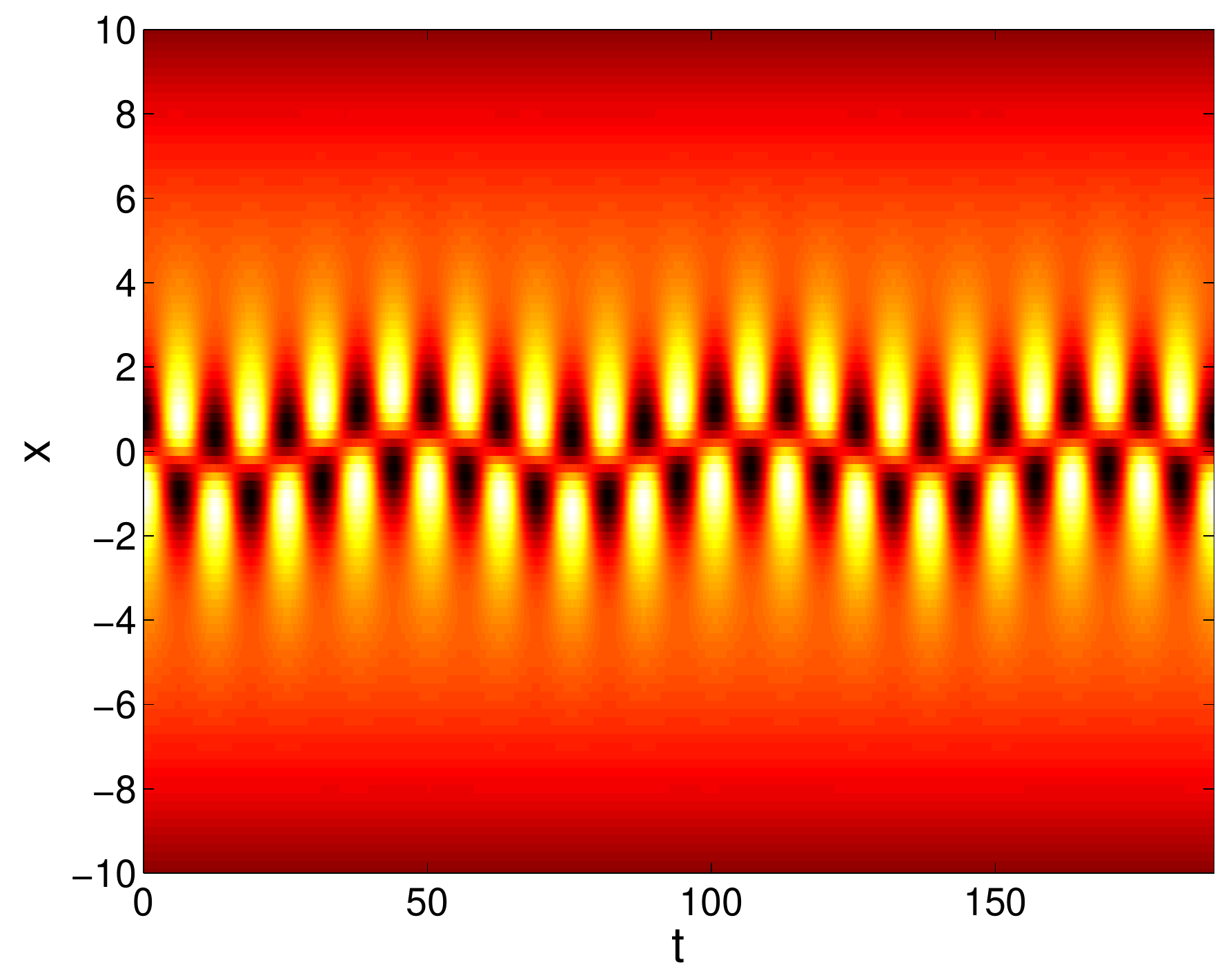} &
 \includegraphics[width=4cm]{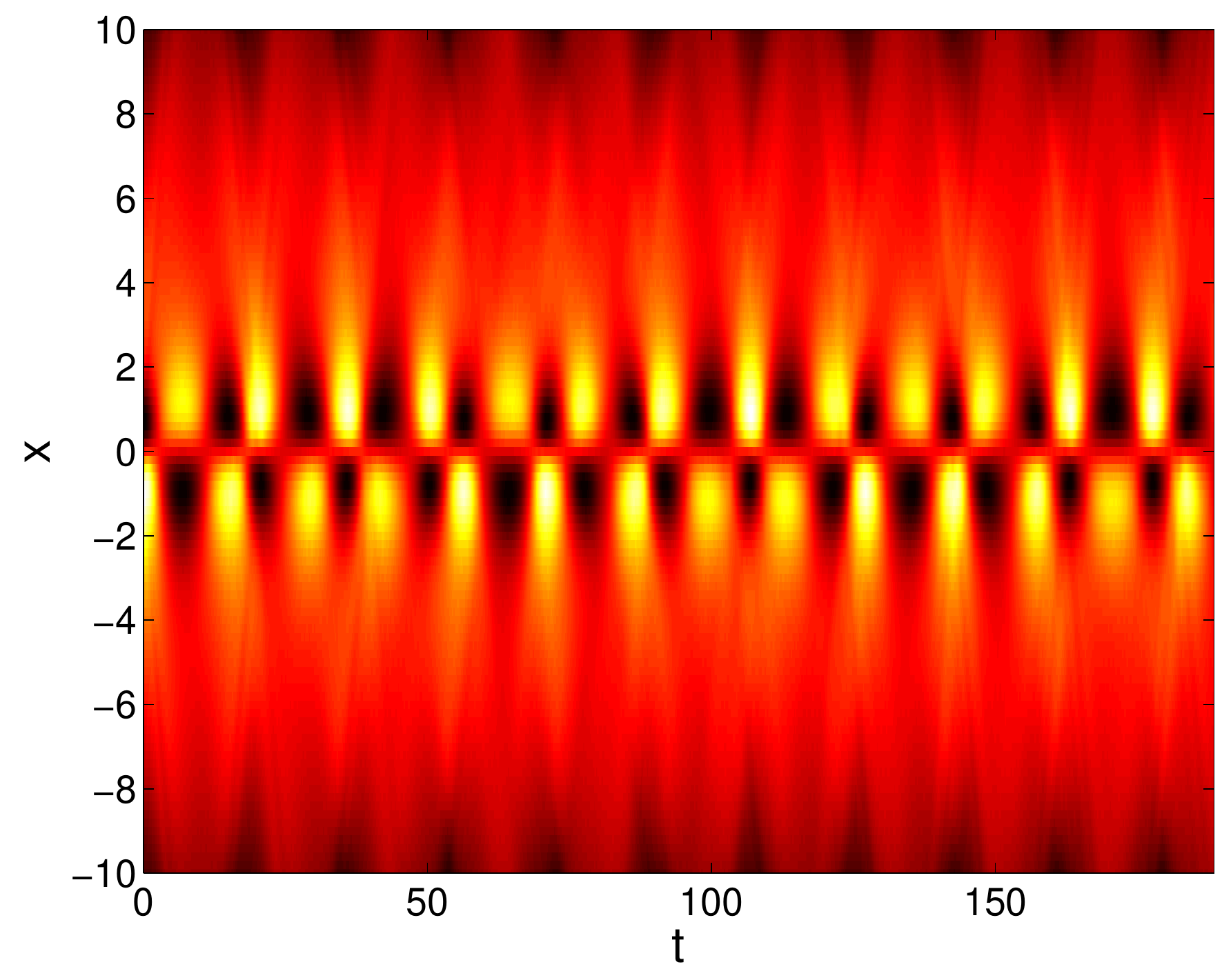} &
 \includegraphics[width=4cm]{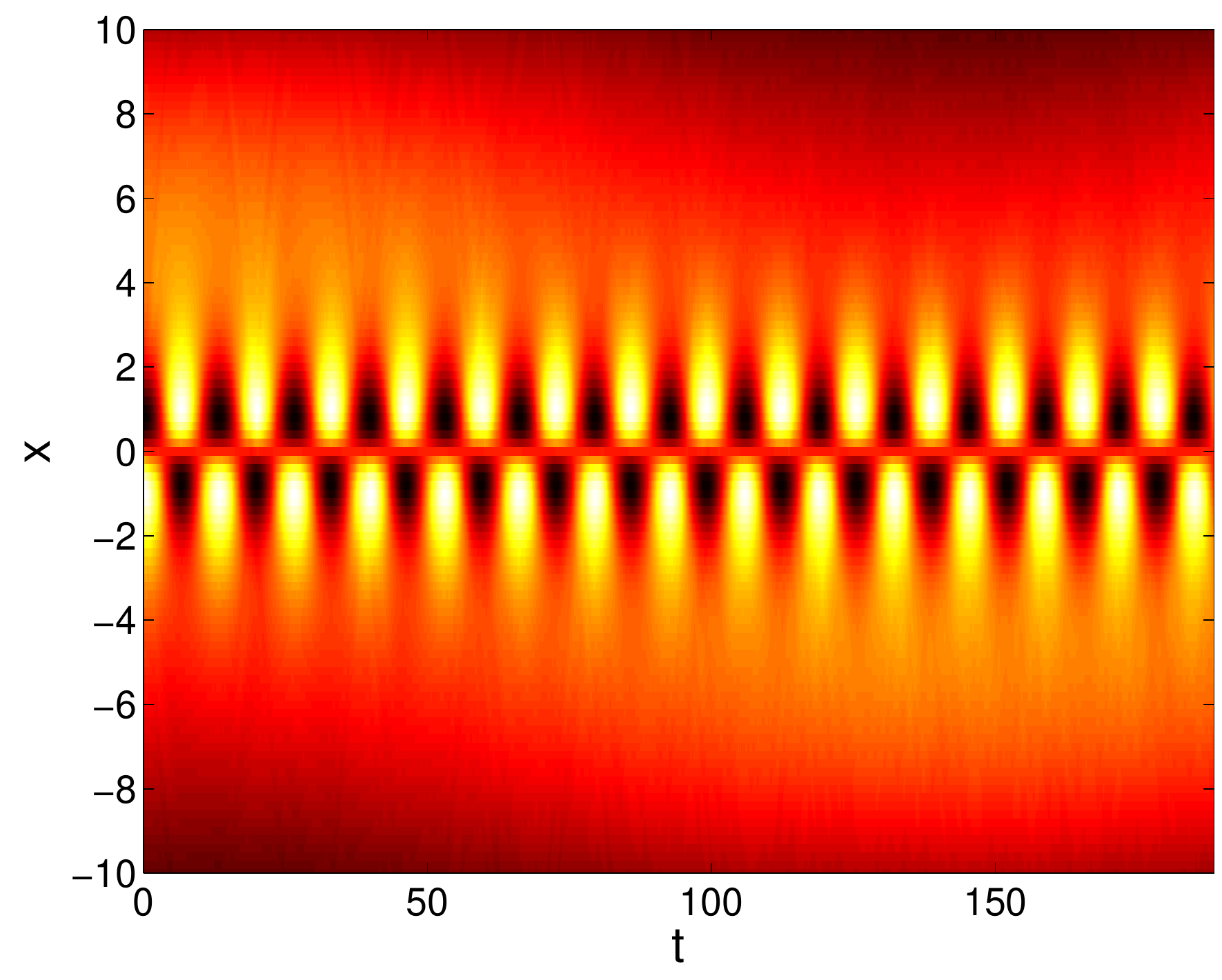} \\
 \includegraphics[width=4cm]{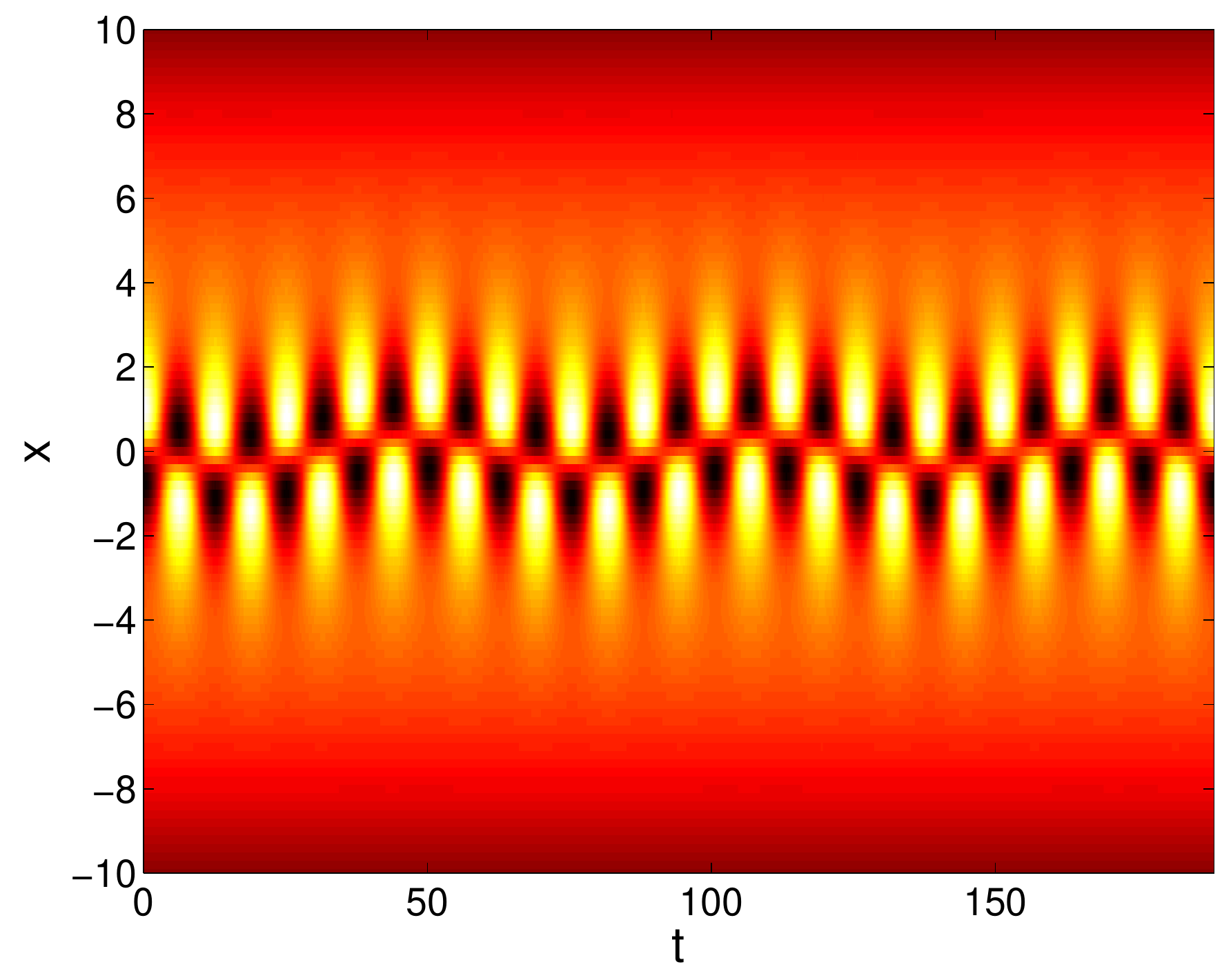} &
 \includegraphics[width=4cm]{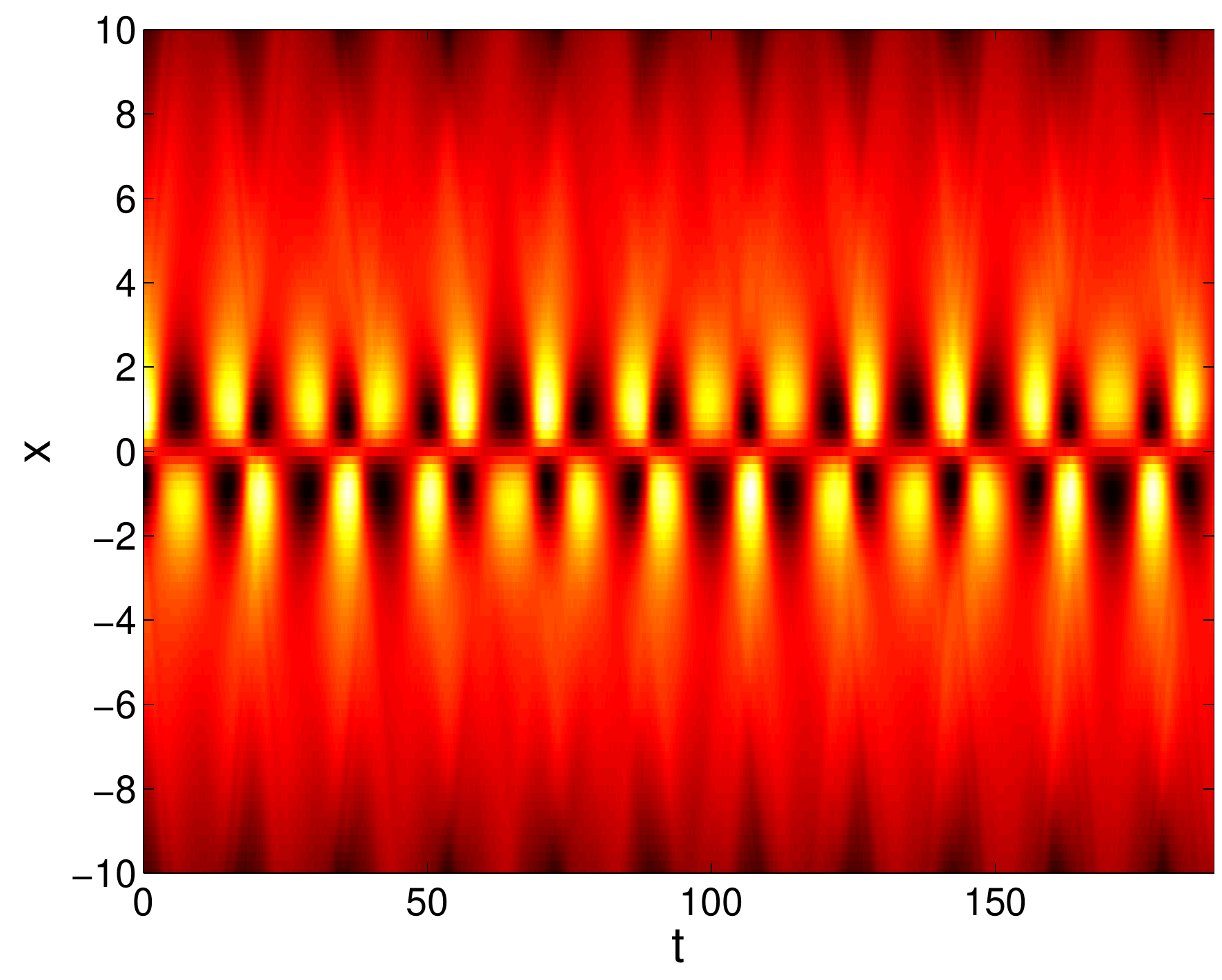} &
 \includegraphics[width=4cm]{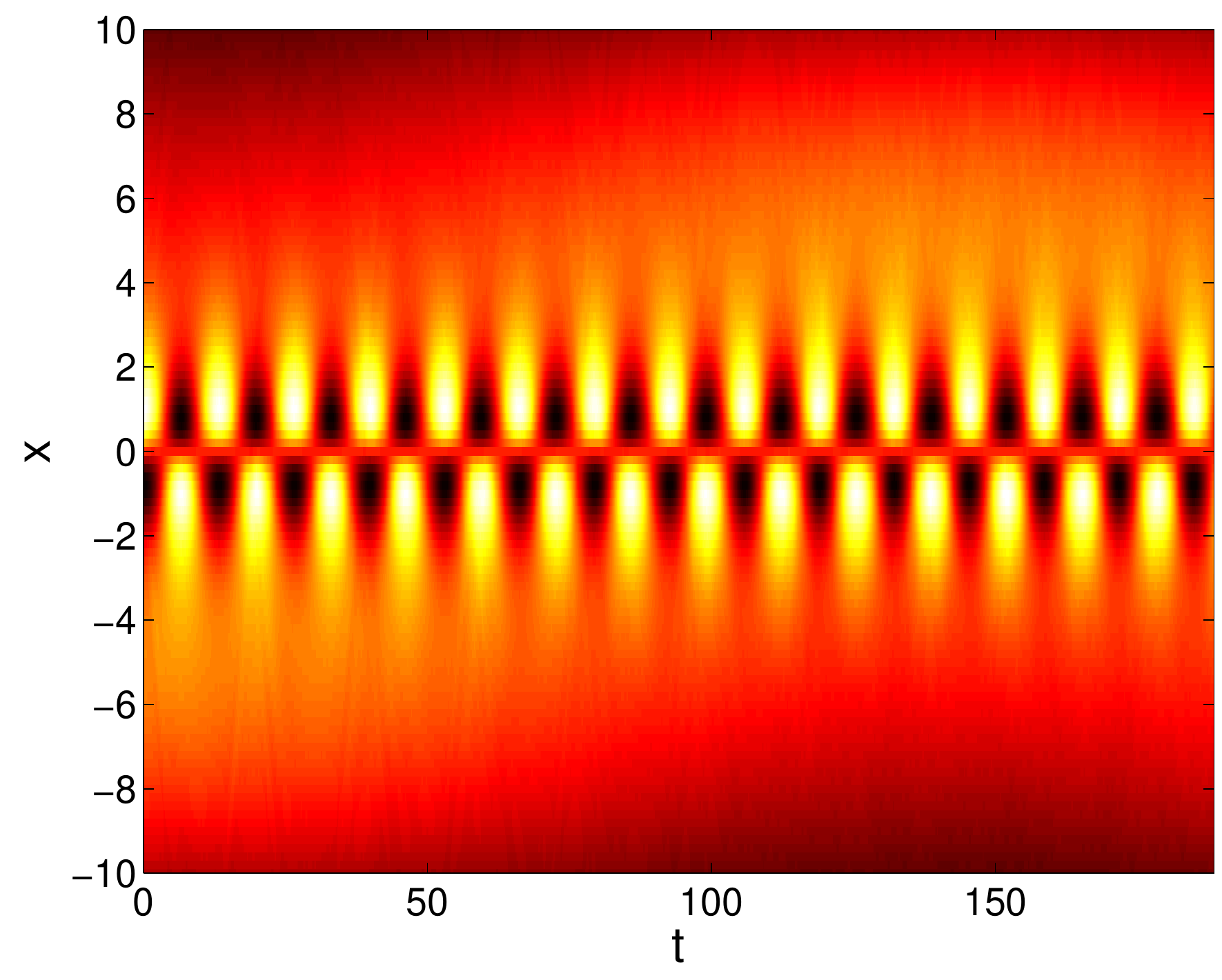} \\
\end{tabular}
\caption{Contour plots showing the evolution of the densities of the
  DD soliton components when perturbed by three different eigenmodes:
  the left panel corresponds to a soliton perturbed along the blue
  mode and leads to trap oscillations; in the central panel, the
  perturbation (along the black mode) leads to a breathing of the
  soliton width, whereas in the right panel (perturbation along the
  green mode), the outcome corresponds to an oscillation of the whole
  condensate. In both cases, $\mu=3$, $\wo=0.5$ and $\Omega=0.2$.}
\label{fig:perturb2}
\end{figure}

Some interesting results can be extracted by the analysis of the
internal modes of the periodic orbits. Figure~\ref{fig:modes1}(left)
shows the dependence of three internal modes of the Floquet spectrum
with respect to $\mu$ for $\wo=0.5$. The blue line is close to the
frequency predicted by Eq.~(\ref{omega_osc}) [depicted as a dashed red
line]. Indeed, perturbing the beating DD soliton with the
corresponding eigenmode, we have confirmed that this perturbation
leads to an oscillation of the soliton in the trap with a frequency
equal to that of the eigenmode (cf. left panel of
Fig.~\ref{fig:perturb2}). It can be observed that the agreement between
the numerical eigenfrequency and that predicted by
Eq.~(\ref{omega_osc}) improves when $\mu$ increases, as expected. The
right panel of Figure~\ref{fig:modes1} shows the dependence of the
frequency of the internal mode corresponding to the oscillation of the
trap with respect to $\wo$ for fixed $\mu=5$ and compares it with the
frequency predicted by Eq.~(\ref{omega_osc}).

We note here, as an aside
%that
in the case $\Omega=0$, that the
  internal soliton modes are neutral modes located at (1,0) on the
unit circle.
In
particular, the mode associated with the oscillation of the DD soliton
in the trap becomes in this case a neutral mode associated with the
translation of the soliton.  The algebraic multiplicity of the
multiplier at $(1,0)$ in the case of $\Omega=0$ is $8$, while in the
trapped case (due to the lifting of translational invariance) it is
$6$.

In order
to observe the properties of other internal modes, we have perturbed
the beating DD soliton with the corresponding eigenmodes. In
particular, a perturbation along the direction of the localized mode
depicted in black in the left panel of Fig. \ref{fig:modes1}, leads to
a breathing in the width of the soliton --- see left panel of
Fig.~\ref{fig:perturb2}. On the other hand, a perturbation along the
direction of the mode depicted in green in the left panel of
Fig.~\ref{fig:modes1}, leads to an oscillation of the condensate along
the trap, leaving the beating DD soliton unaffected (i.e., the soliton
stays at the trap center) --- see right panel of
Fig.~\ref{fig:perturb2}.

\begin{figure}
\begin{tabular}{cc}
 \includegraphics[width=6cm]{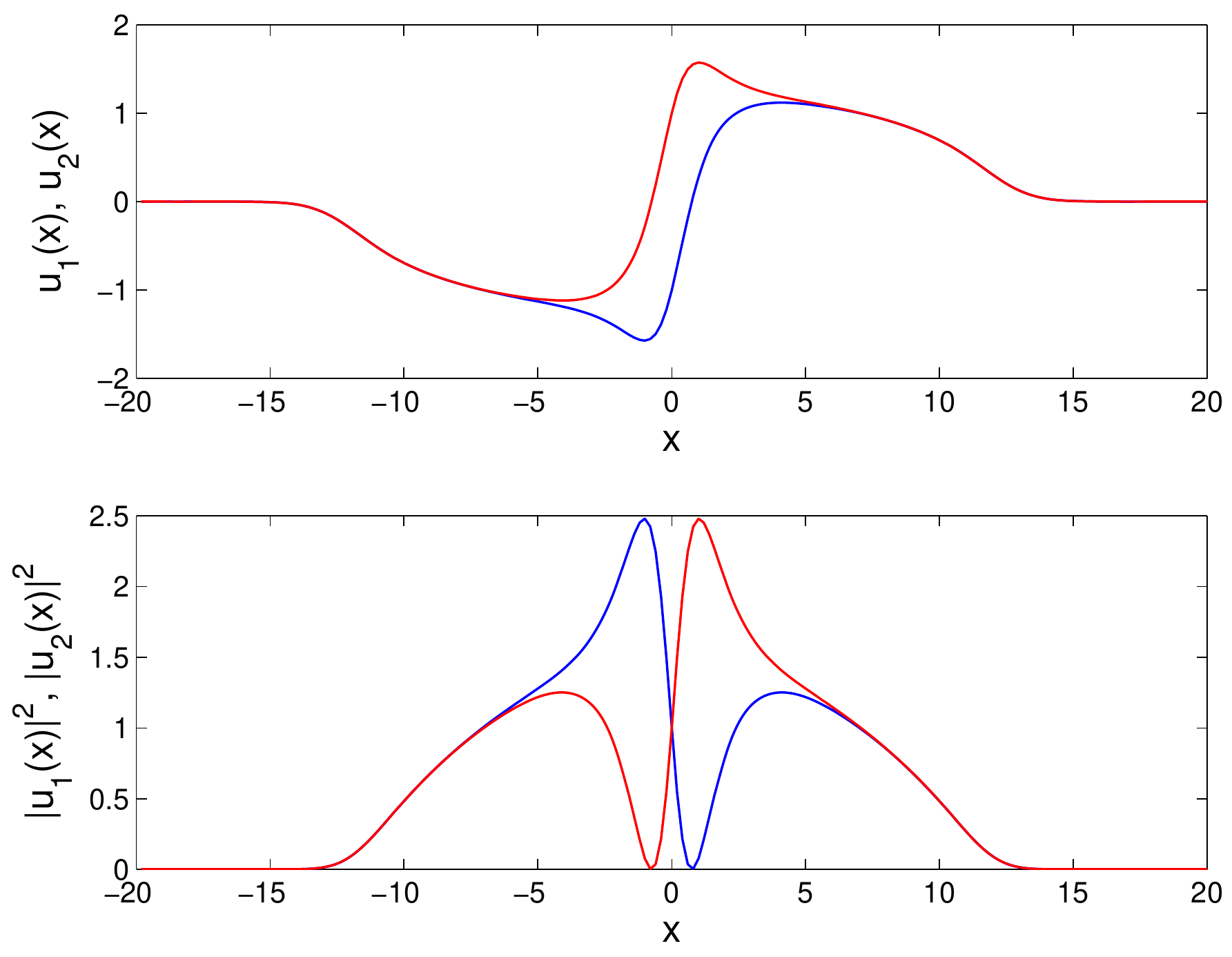} &
 \includegraphics[width=6cm]{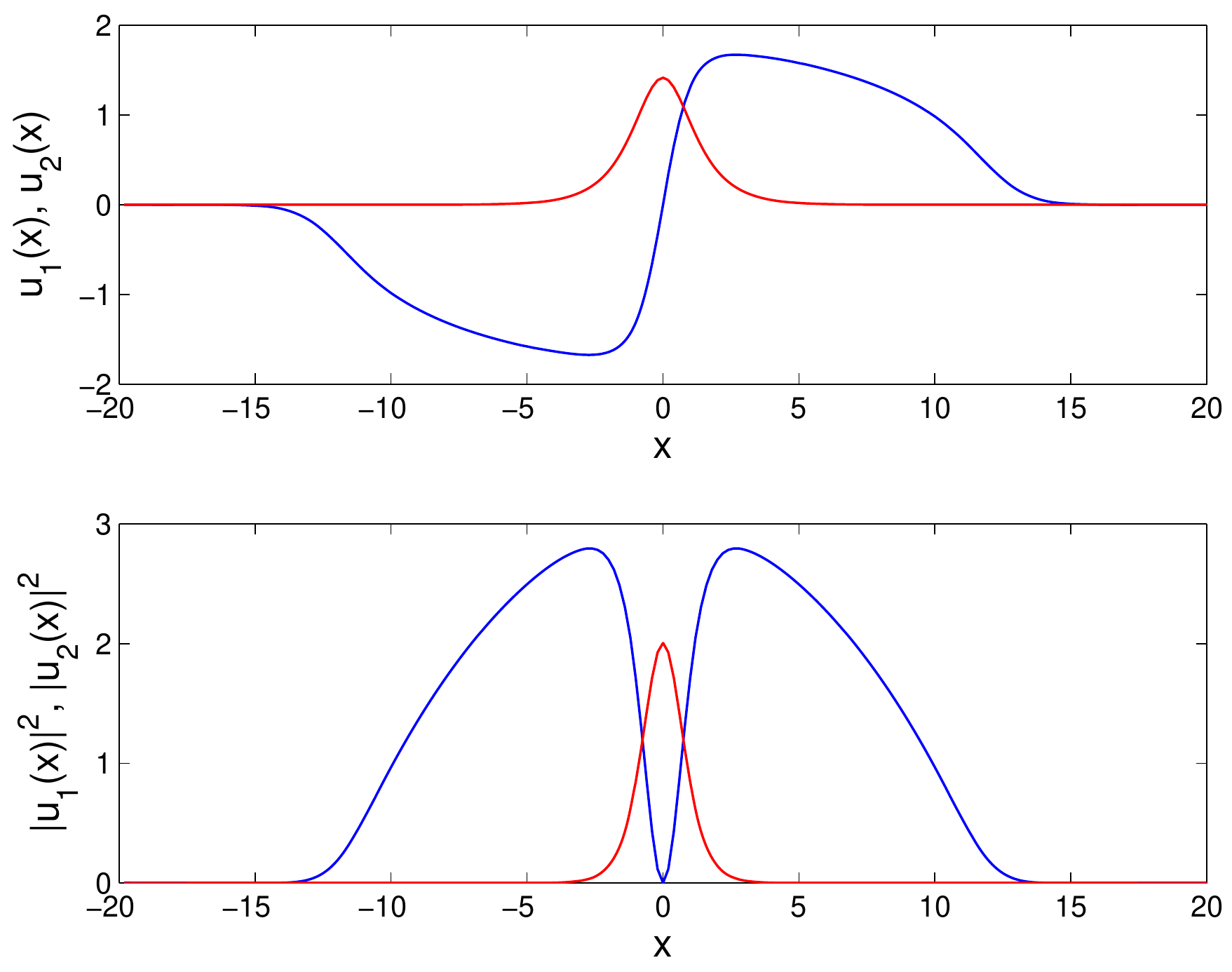} \\
\end{tabular}
\caption{(Left panel) Profiles and densities of a periodic orbit at $t=0$ with $\mu=3$, $\wo=0.5$ and $\Omega=0.2$. The right
panel shows the dark-bright soliton arising by rotating with $\chi=-\pi/4$ the dark-dark soliton of the left panel.}
\label{fig:example}
\end{figure}

Finally, we make a remark about the way we have calculated the value
$N_B$ that must be introduced in Eq.~(\ref{omega_osc}). The procedure
consists in performing an SO(2) rotation with $\chi=-\pi/4$ to the
periodic DD soliton at $t=0$. This solution is shown in the left panel
of Fig. \ref{fig:example}, whereas the rotated solution is depicted in
the right panel of the same figure. Thus, $N_B$ is the norm of the
bright component of the rotated solution. It can also be inferred from
the Fourier coefficients of the periodic orbit:

\begin{equation}
    N_B=\int_{\mathbb{R}}|u_2|^2\mathrm{d}x=\frac{1}{2}\int_{\mathbb{R}}
    [|\sum_k z_k|^2+|\sum_k y_k|^2-2\mathrm{Re}((\sum_k z_k^*)(\sum_k y_k))]\mathrm{d}x.
\end{equation}

\section{Single beating dark-dark solitons near and far from the
  Manakov limit}
%states with and without a trap:
%numerical results}

\begin{figure}[t]
 \includegraphics[width=8cm,height=6.2cm]{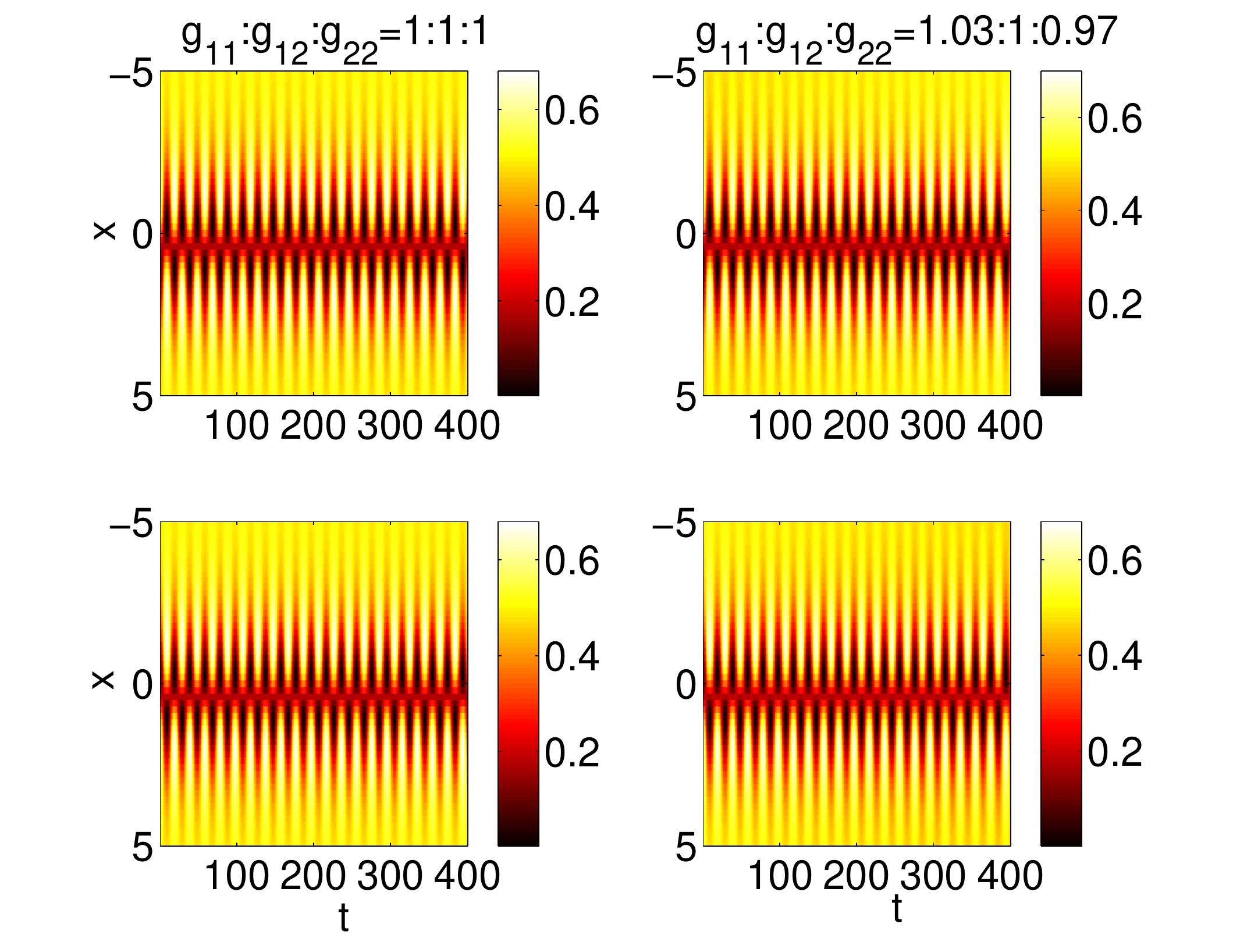}
\caption{The comparison between the integrable case
$g_{11}:g_{12}:g_{22}=1:1:1$ (left column) and the non-integrable
case $g_{11}:g_{12}:g_{22}=1.03:1:0.97$ (right column),
%which are demonstrated in the left and the right column
%respectively.
is demonstrated.
The upper panels show the densities of the first dark component while the
lower ones show
the second dark component. Here $\eta=0.6, \chi=\pi/4, \theta=0, k=0, \mu=1$.
Based on the similarity of the relevant dynamics, we will focus
on the case of unit nonlinear coefficients.}
\label{compare_g11_g22_nontrapped}
\end{figure}

We now turn to a numerical study
%examination
of the
%numerical
properties of
%such
the beating DD soliton states.  Firstly, in the absence of a trap, we
are going to compare the integrable case with equal scattering lengths
$g_{11}:g_{12}:g_{22}=1:1:1$ to the non-integrable case
$g_{11}:g_{12}:g_{22}=1.03:1:0.97$ (see Ref.~\cite{mertes}). From
Fig.~\ref{compare_g11_g22_nontrapped}, we observe that both of the
dark components are oscillating with fixed frequencies and these two
cases are very similar~\footnote{In what follows when the relevant
interaction coefficients are not explicitly mentioned, it will be implied that
they assume the values  $g_{11}=g_{12}=g_{22}=1$.}
(a feature that we have generically observed
between the dynamical phenomenology of these two cases).

\begin{figure}[t]
\includegraphics[width=8cm,height=6.2cm]{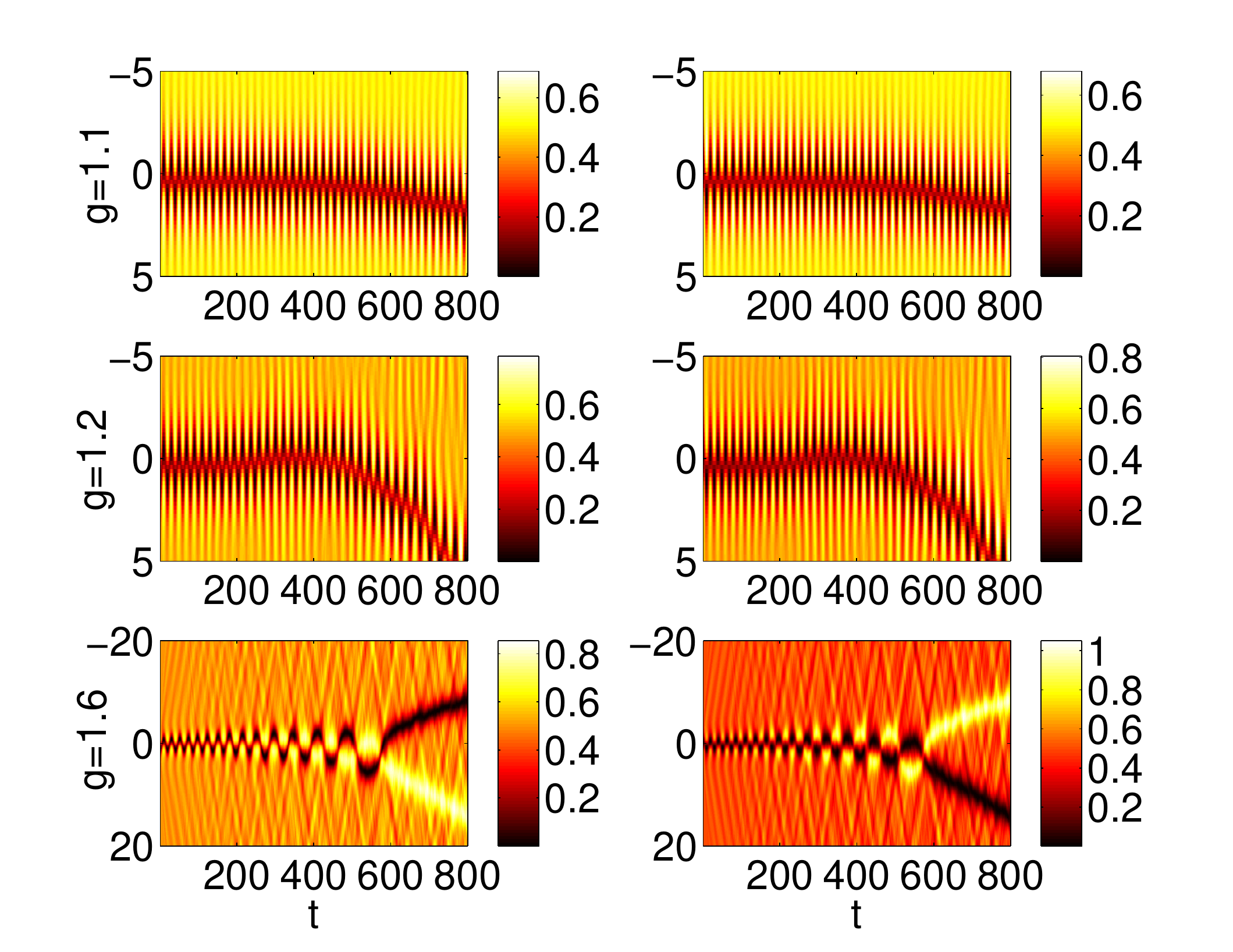}
\caption{The dynamics of the densities of a DD soliton in the absence of
a trap but for $g=1.1, g=1.2, g=1.6$ respectively ($g \equiv g_{11}$).
When $g=1.1$ or $g=1.2$, the soliton is set into translational motion.
However, for $g=1.6$  the coherent structure executes a growing
oscillation which eventually results in its splitting into a pair of
dark-antidark
%solitary waves
solitons (i.e., a dark soliton in the one component
coupled to a lump in the other).
The parameters used here are the same as
in Fig.~\ref{compare_g11_g22_nontrapped}.}
\label{nonintegrable_compare}
\end{figure}

To highlight the fact that substantial variations of the scattering
length--which can be imposed by virtue of a Feshbach resonance--may
have a significant impact on the robustness of the beating DD
solitons,
%state,
we consider scattering lengths in the set with ratios
$g_{11}:g_{12}:g_{22}=g:1:1$.  In particular, we take $g=1.1, 1.2,
1.6$ in Fig.~\ref{nonintegrable_compare}. When $g$ is not so large,
i.e., $g=1.1, 1.2$, the beating DD
%dark-dark
soliton oscillates and, as $t$ increases, the change in the scattering
length results in mobility of the coherent structure.
%it is shifted and the equilibrium of the oscillation is
%changed.
However, more dramatic events can arise when $g$ is relatively large,
e.g., for $g=1.6$. There, we can see that the soliton is finally split
into two fragments (upon growth of the intrinsic beating oscillation
which eventually induces the splitting) and results in two states that
resemble dark-antidark solitons~\cite{epjd} (see also
Ref.~\cite{hadi}).  In particular, each of the components acquires a
dark soliton coupled to a lump in the second component.

%

\begin{figure}
 \includegraphics[width=8cm,height=6.2cm]{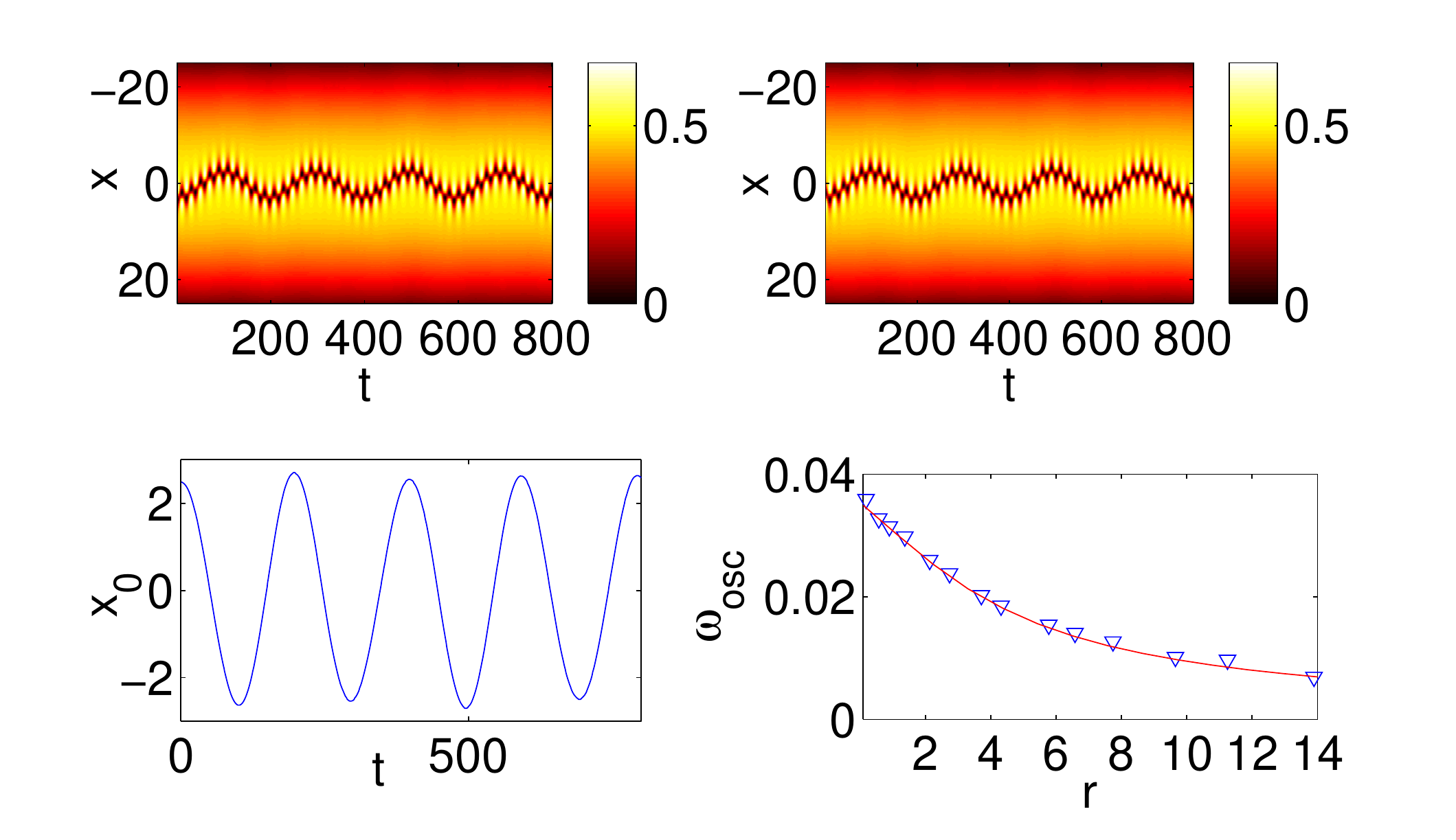}
\caption{Motion of a dark-dark soliton in
%within
a trap of strength
%with trap frequency
$\Omega=0.05$. Other parameters
are the same as in Fig.~\ref{compare_g11_g22_nontrapped}. The upper two
panels show the oscillation
of the wave initially centered at $x_0=2.5$ (the chemical potential is
$\mu=1$).
The lower left panel demonstrates the center of mass of the DD in the upper
panels. The analytical
oscillation frequency, given by Eq.~(\ref{omega_osc}), is 0.03123, while the
numerical frequency, calculated by Fourier transform, is 0.03238. The
lower right panel yields the comparison between
the analytically calculated frequencies (red line)
versus the numerical obtained ones (the blue triangles), as
$r$ varies between 0.1 and 14.}
\label{frequency_trap}
\end{figure}

% Next, let us consider
% %Now, we are about to study
% the single beating DD
% %one dark-dark
% soliton solution within a harmonic trap of the form
% %i.e., for
% $V(x)=\frac{1}{2}\Omega^2x^2$. In the presence of the trap, the dynamics
% of the center
% of mass $X_0(t)$ of the beating DD soliton is still described by the dynamics
% of the original
% (unrotated) DB soliton center $x_0$. The latter is known to perform harmonic
% oscillation in the trap according to the equation
% $\ddot{x_0}+\omega_{osc}^2 x_0=0$, where the oscillation frequency
% $\omega_{osc}$ is given by \cite{BA}:
% %
% \begin{eqnarray}
% \omega_{osc}^2 &=& \Omega^2\left(\frac{1}{2}-\frac{r}{8\sqrt{1+(\frac{r}{4})^2}}\right)
% \label{omega_osc}
% \end{eqnarray}
% %
% where $r=\frac{N_B}{\sqrt{\mu}}$ is a measure of the ratio of number
% of atoms in the bright and dark soliton component.
In Fig.~\ref{frequency_trap}, we show a particular
example of the DD soliton
%within a
in the trap,
%Actually, it is oscillating around equilibrium points
%which are shown in
%upper two panels,
which oscillates around its center; the parameter values are
%for
$\mu=1$, $\eta=0.6$, initial soliton position $x_0(t=0)=2.5$, and trap
strength $\Omega=0.05$. Note that for these runs, the initial profile
of the beating DD soliton in the trap is approximated by the
numerically found (in trap) ground state --- i.e., the Thomas-Fermi
cloud --- multiplied by the beating DD solution (without a trap) of
Eqs.~(\ref{new_u1_sim})-(\ref{new_u2_sim}). Then via a
%our
time-stepping algorithm (a fourth-order Runge-Kutta scheme),
we obtain the time evolution of the densities of the oscillating solitons
%of oscillation
in the upper two panels. Moreover,
the left lower panel shows the center of mass of the beating DD soliton
%within a trap of trap frequency
in the trap. Using Fourier analysis, we can infer
the numerical frequency of in-trap oscillation, which
can, in turn,
be compared to the analytical one,
%computed by
cf. Eq.~(\ref{omega_osc}).
As shown in the bottom right panel of the figure, there is very good
agreement between the two.
%Finally, we plot the analytical and numerical frequency of the DD oscillation
%versus the ration $r$. Obviously, the Eq.~(\ref{omega_osc}) can predict the frequency quite
%well.

%
\begin{figure}
 \includegraphics[width=8cm,height=6.2cm]{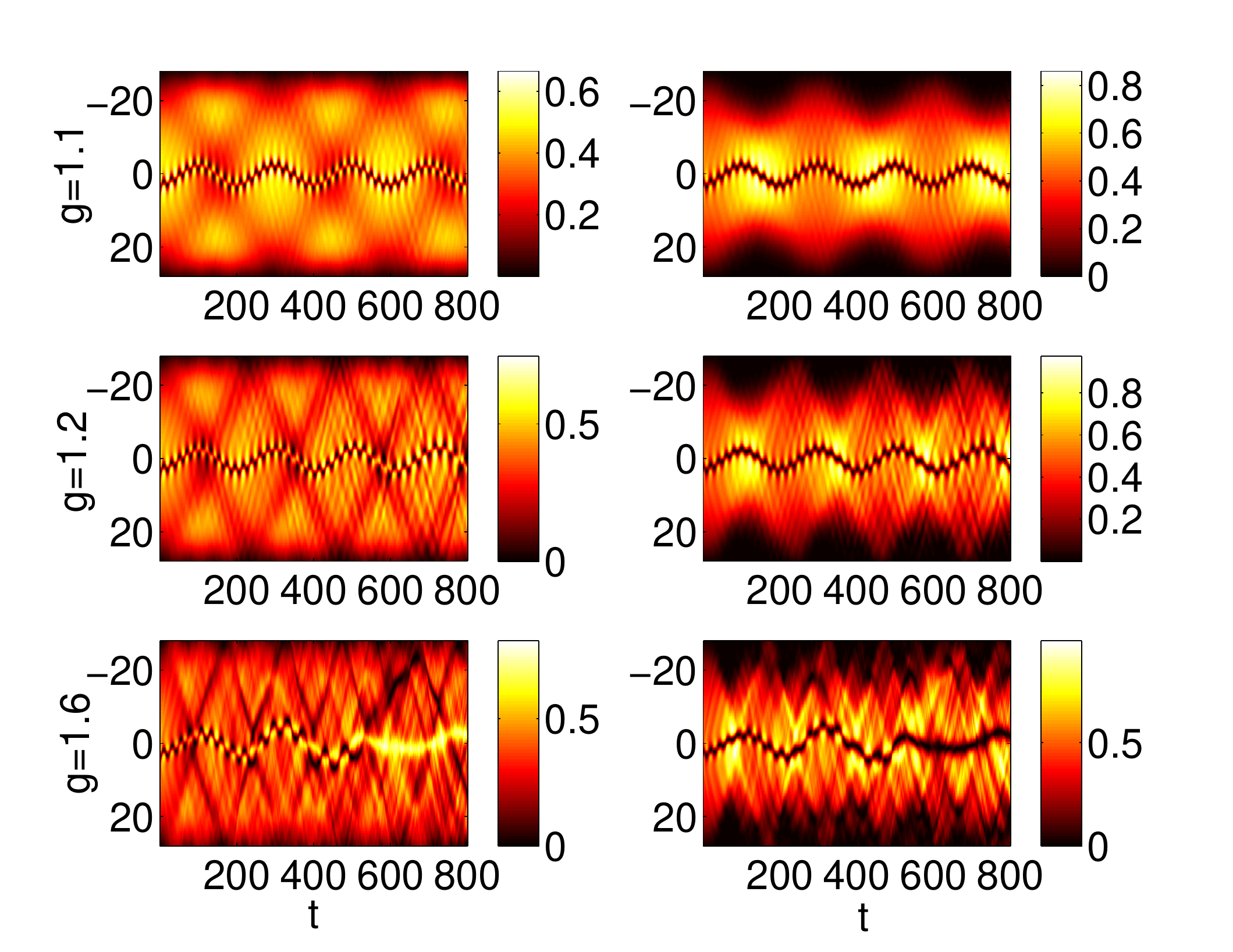}
\caption{The comparison of the oscillation of the
density of a DD soliton within a
trap of trap frequency
$\Omega=0.05$ for different values of $g$.
The soliton is initialized at $x_0=2.5$; g is set to be
$1.1$ (top panels), $1.2$ (middle panels), $1.6$ (bottom panels)
for the combination
of the scattering length $g_{11}:g_{12}:g_{22}=g:1:1$.
Other parameters are similar to Fig.~\ref{frequency_trap}.}
\label{g_11_12_16_compare_trap}
\end{figure}

Next, we consider the in-trap dynamics of a single beating DD soliton
but for the non-integrable cases. Again, when
$g_{11}:g_{12}:g_{22}=1.03:1:0.97$, we observe a nearly identical
phenomenology to that of unit $g_{ij}$'s.  For the more significant
deviations from that case of the form $g_{11}:g_{12}:g_{22}=g:1:1$
where $g=1.1, 1.2, 1.6$, the results are reported in
Fig.~\ref{g_11_12_16_compare_trap}. For lower values of $g=1.1, 1.2$,
the behavior of the DD is similar to the case with $g=1$, however, we
progressively observe more significant radiative emissions which also
affect the oscillation frequency. However, once again the
modifications of the phenomenology are most dramatic in the case of
$g=1.6$ of the bottom panels. There, the radiation emission is
accompanied by growing intrinsic oscillations which eventually result
in the breakup and formation of a single dark-antidark solitary wave.

\section{Two dark-dark soliton states: dynamics and interactions}
%Finally,
We now consider the interactions of two beating
%dark-dark
DD solitons.
%in the GPE model.
%in BEC and the model equations used here are
%Eqs.~(\ref{deq1})-(\ref{deq2}). And first of all,
We once again start from the untrapped case and use
%Using again as our starting point,
%the DB waveforms, we can use
as an initial ansatz a two-DB soliton state of the form:
%As we know,
%the two dark-bright solition solution has the form:
%
\begin{eqnarray}
u_1 &=& \left(\cos\phi\tanh\xi_{-}+i
  \sin\phi\right)\left(\cos\phi\tanh\xi_{+}-i \sin\phi\right)
\label{u1_two_db}
\\
u_2 &=& \eta\sech
\xi_{-}e^{i\left(kx+\theta(t)\right)}+e^{i\Delta\theta}\eta
\sech\xi_{+}e^{i\left(-kx+\theta(t)\right)}
\label{u2_two_db}
\end{eqnarray}
where $\xi_{\pm}=D(x\pm x_0)$, $2x_0$ is the relative distance between the two solitons, and
$\Delta\theta$ is the relative phase between the two bright solitons.
Below we consider both
the out-of-phase (OOP) case, $\Delta\theta=\pi$, as well as the in-phase (IP) case
%corresponding to
$\Delta\theta=0$. Once again taking advantage of the model invariance
under the SO(2) rotations, as we did for the single DD soliton case, we use the orthogonal matrix (\ref{so2})
%as SO(2) rotations as follows:
%
%\[ U=\left( \begin{array}{ccc}
%\cos(\chi) & -\sin(\chi) \\
%\sin(\chi) & \cos(\chi) \end{array} \right)\]
%
%where $\chi$ is an arbitray angle. We can thus
and obtain a two-beating-DD-soliton
{\it ansatz} in the form:
\begin{eqnarray}
u_1 &=& \cos(\chi)\left(\cos\phi\tanh\xi_{-}+i
  \sin\phi\right)\left(\cos\phi\tanh\xi_{+}-i \sin\phi\right)
\nonumber\\
&-&\sin(\chi)\left(\eta\sech
  \xi_{-}e^{i\left(kx+\theta(t)\right)}+e^{i\Delta\theta}\eta
  \sech\xi_{+}e^{i\left(-kx+\theta(t)\right)}\right),
\label{u1_new_two_db}
\\
u_2 &=& \sin(\chi)\left(\cos\phi\tanh\xi_{-}+i
  \sin\phi\right)\left(\cos\phi\tanh\xi_{+}-i \sin\phi\right)
\nonumber\\
&+&\cos(\chi)\left(\eta \sech
  \xi_{-}e^{i\left(kx+\theta(t)\right)}+e^{i\Delta\theta}\eta
  \sech\xi_{+}e^{i\left(-kx+\theta(t)\right)}\right).
\label{u2_new_two_db}
\end{eqnarray}

\begin{figure}
 \includegraphics[width=8cm,height=6.2cm]{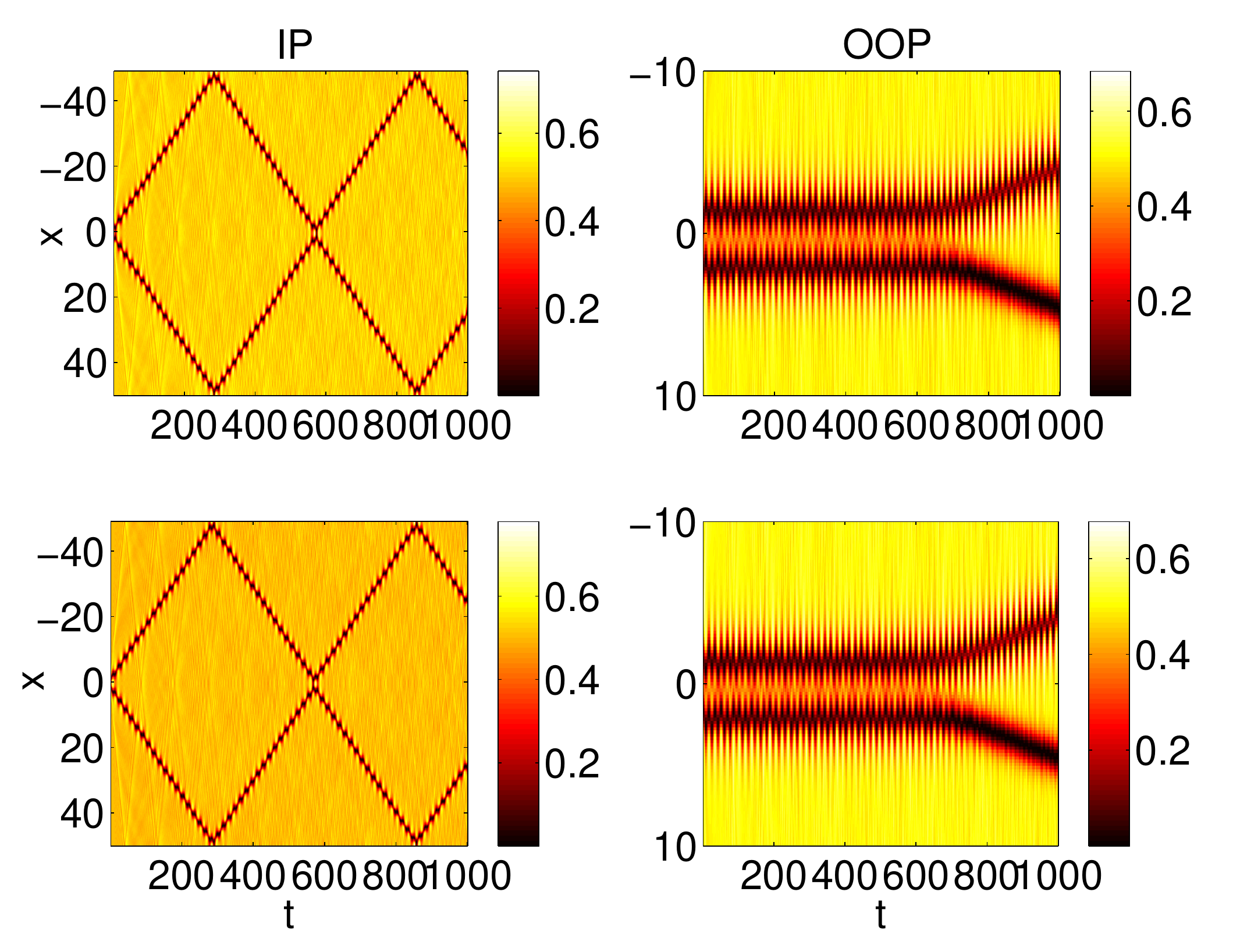}
\caption{Space-time contour plots of two beating DD soliton densities
in phase (left) and out of phase
(right) for $g_{11}=g_{12}=g_{22}=1$ .
Here $\chi=\pi/4, \eta=0.5, x_0=1.5, D=1.2$.}
\label{compare_in_out_phase_untrapped}
\end{figure}

In our numerical study for the dynamics of the two-beating-DD-soliton state, we
%We
first consider
%the two DD soliton initial condition in
the integrable case, corresponding to $g_{11}=g_{12}=g_{22}=1$, both
for the in-phase and out-of-phase cases. The results of the
simulations, using initial conditions corresponding to the above
ansatz, are shown in Fig.~\ref{compare_in_out_phase_untrapped}.
%, using the above initial guesses.
In the in-phase case, the repulsion between the beating DD solitons is
immediately evident resulting in the strong separation of the two
waves (which still perform their internal beating).  On the other
hand, in the out-of-phase case, the competition between the repulsion
of the dark components and the attraction between the bright
components of the progenitor DB solitons (see Ref.~\cite{multiarxiv})
can be discerned, as the configuration remains nearly stationary
for a lengthy evolution interval. Finally, however, the repulsive
interaction prevails and the solitons eventually separate.

%we use the Eqs.~(\ref{u1_new_two_db})-(\ref{u2_new_two_db}) as the initial guess of the two DD
%and do the integration by the fourth-order Runge-kutta, where $\Delta\theta=0, \pi$ for the in phase
%and out of phase respectively.
%For the in phase case, the two DD repel with each other and still
%oscillate around the equilibrium. For the out of phase, they do not repel at the beginning, but
%as time goes by, they repel and do not oscillate near the equilibrium.
%

\begin{figure}
 \includegraphics[width=8cm,height=6.2cm]{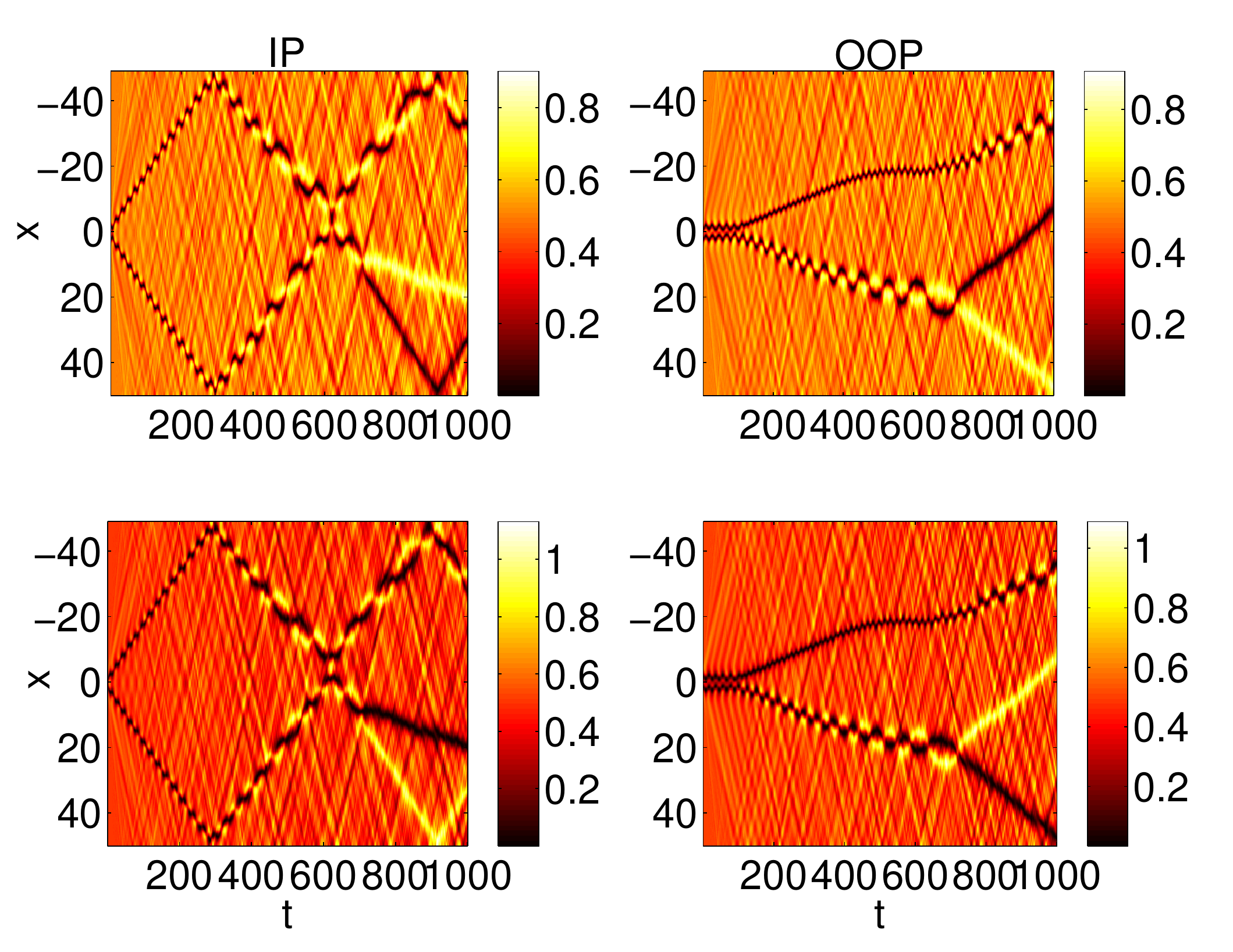}
\caption{Space-time contour plots of two DD soliton densities in
phase (left) and out of phase
(right) for $g=1.6$ in the set $g_{11}:g_{12}:g_{22}=1.6:1:1$.
Here $\chi=\pi/4, \eta=0.5,
x_0=1.5, D=1.2$.}
\label{g_16_compare_untrapped}
\end{figure}

Next, we consider the non-integrable case. Since for
$g_{11}:g_{12}:g_{22}=1.03:1:0.97$, the phenomenology is again very
similar to $g_{11}=g_{12}=g_{22}$, we consider the significant departure
from this limit pertaining to $g_{11}:g_{12}:g_{22}=1.6:1:1$.
In Fig.~\ref{g_16_compare_untrapped}, we observe
that in the in-phase
%(IP)
case, the two beating DD solitons initially separate and move away from each other,
%are %strongly separated although highly nontrivial events follow their
then they are reflected
%return
from the domain boundary and a new collision occurs.
%recombination).
After this collision, a highly nontrivial event is observed, namely one of the two beating DD solitons
is decomposed into a dark-antidark soliton pair, with each of these solitons moving with different velocities.
For the out-of-phase
%(OOP)
case, the separation arises much faster than
for the unit coefficients and, interestingly, results in an asymmetric evolution
with one of the DD solitons breaking up in a pair of dark-antidark solitons
%solitary waves
(as in Fig.~\ref{nonintegrable_compare} of section II).
%we saw in the previous section), while
Notice that, as in the in-phase case, the other soliton
%one has
is not broken up in a similar way during the horizon of the simulation although it is
likely that such an event will also occur for that wave.

\begin{figure}
\includegraphics[width=8cm,height=6.2cm]{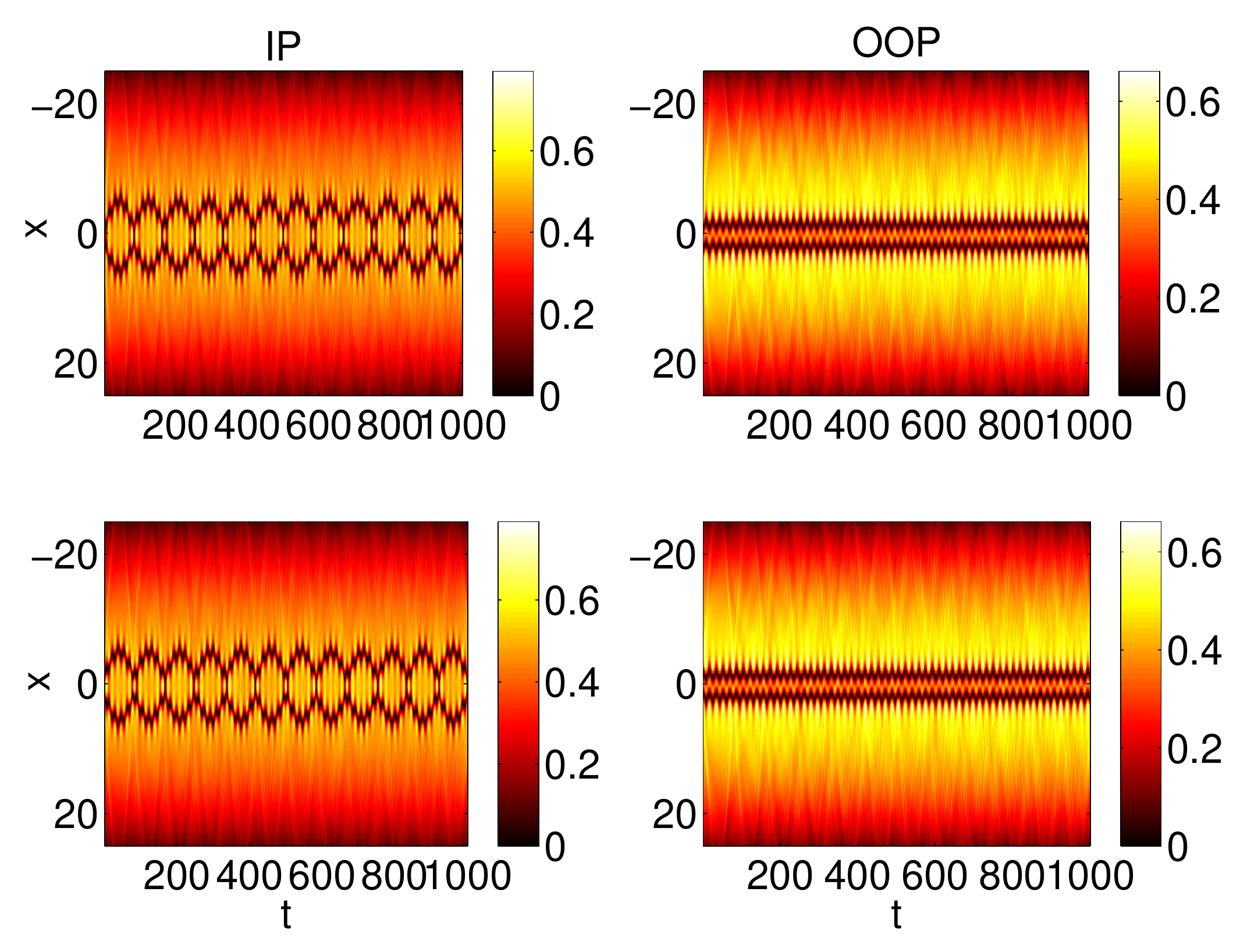}
\caption{Space-time contour plots of two beating DD soliton densities in phase (left)
and out of phase
(right) in the case of equal $g_{ij}$'s
within a harmonic trap with trap frequency $\Omega=0.05$. Here
$\chi=\pi/4, \eta=0.5, x_0=1.5, D=1.2, \mu=1$.}
\label{compare_in_out_phase_trapped}
\end{figure}

Next, we consider the two-beating-DD soliton in the trap, in the case of unit coefficients.
%IP and OOP but in the trap.
We set $V(x)=\frac{1}{2}\Omega^2x^2$,
%where
with $\Omega=0.05$, and the chemical potential
$\mu=1$. From Fig.~\ref{compare_in_out_phase_trapped},
we infer that the two beating DD solitons are now trapped and oscillate
around an equilibrium position. Notice that in the in-phase case, the solitons perform out-of-phase oscillations and
undergo quasi-elastic collisions.
while
%In fact, in the OOP
In the out-of-phase case,
%with
the weak residual repulsion is counter balanced by the
%of the case without the trap having been countered by the
presence of the trap, and we observe that
%a near-immobility of
the two beating DD solitons remain in a close distance to each other.
%solitary waves in the dynamics.

Finally, we consider two DD with $g_{11}:g_{12}:g_{22}=1.6:1:1$
within the same trap.
%as the integrable case, then to see what the dynamics of two DD will be.
%Just as the Fig.~(\ref{g_16_trapped}) shows that, if the nonlinear coefficient is large i.e.
%$g=1.6$, then a lot of radiations come out in the IP, and the DD do not oscillate in the OOP.
In this case, we observe that despite the presence of the trap, it is
not possible to sustain a robust set of oscillations and interactions
between the beating DD solitons.
%dark-dark solitary waves.
This is especially true in the out-of-phase
%OOP
case, where the oscillatory growth of the beating eventually
results in the breakup of the DD soliton states into dark-antidark ones
(which generally appear more robust for such higher values of
$g$).
\begin{figure}
 \includegraphics[width=8cm,height=6.2cm]{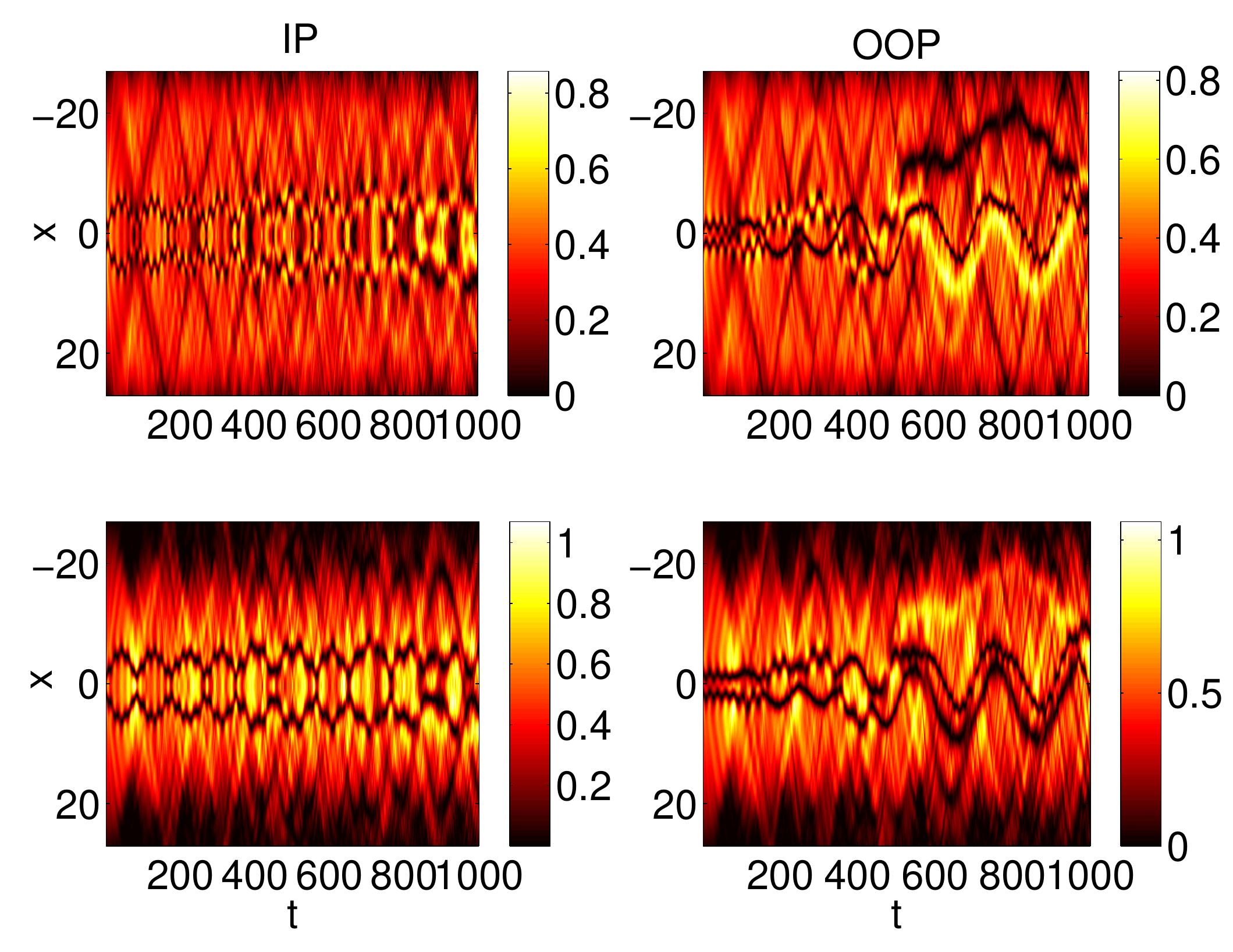}
\caption{Space-time contour plots of two beating DD soliton densities in phase (left)
and out of phase
(right) for the case with $g_{11}:g_{12}:g_{22}=1.6:1:1$
within a harmonic trap with trap frequency
$\Omega=0.05$. The parameters used are the same as for the previous figure.}
%Here $\chi=\pi/4, \eta=0.5, x_0=1.5, D=1.2, \mu=1$.}
\label{g_16_trapped}
\end{figure}

\section{Conclusions and Future Challenges}

In this work, we have studied the stability and dynamics of beating
dark-dark (DD) solitons in pseudo-spinor
% considered the possibility of multi-component
Bose-Einstein condensates,
%to sustain a novel beating state in the form of a dark-dark solitary wave,
motivated by recent experiments where such structures were observed.
%in this field.
We have illustrated the connection of these solitons
%solitary waves
with internal density oscillations
to dark-bright (DB) solitons identified earlier, through
SO(2) (and more generally SU(2)) rotations. We have illustrated that
such states persist in the presence of the trap and, in fact,
oscillate with the frequency previously predicted for
%the oscillation of
dark-bright solitons.
%in this type of trap),
Using Floquet analysis, we have also identified beating dark-dark
solitons as stable periodic orbits in the integrable (Manakov) limit
with and without a trap.

We have also investigated in detail the effect of the deviation from
the Manakov case by considering different from unity
%as well as for
scattering length ratios.
%are away different from the integrable limit of unit ratios in the $g_{ij}$'s with only weak modifications.
We have shown that when the deviation from the integrable case is
small (as is the physically relevant case of a pseudo-spinor
condensate composed by different spin states of rubidium), then the
stability and dynamics of beating dark-dark solitons follow that of
the integrable case.  However, we also illustrated that a significant
departure of the ratios of the scattering lengths
%$g_{ij}$'s
from this limit (towards the miscible regime) will eventually break up
%these solitary waves
beating dark-dark solitons in favor of dark-antidark soliton
entities. We have also considered the interaction of beating dark-dark
solitons
%such solitary waves
finding a typically repulsive dynamical behavior, which can be
attenuated only in the case where the bright components (of the
progenitor dark-bright solitons, used to create the dark-dark ones)
are out-of-phase (and, hence, attracting each other). In that case,
especially in the presence of a trap, a robust set of
multi-beating-dark-dark-soliton states can be created.

The discussion of DD solitons in this work has focused upon
  those states that can be constructed, in the spatially extended,
  Manakov case, from the SU(2) rotation of a DB soliton and confined
  states in the presence of a trapping potential.  In both cases, each
  component of the DD soliton exhibits the same background flow
  velocity.  In a series of experiments
  \cite{engels1,engels2,engels3,multiarxiv}, a relative flow between
  two condensate components induced by a magnetic field gradient led
  to DB solitons and counterflow-induced modulational instability
  resulting in the formation of a number of beating DD solitons.  It
  is natural, then, to inquire into the effect that relative motion
  between two condensate components has on localized structures.  In
  the integrable case, the most general DD soliton was constructed
  using a B\"{a}cklund transformation \cite{parkshin}.  Because it
  allows for a counterflow, this soliton is characterized by eight
  free parameters in contrast to the seven parameter SU(2) rotated DB
  soliton studied here or the seven parameter static DD soliton
  \cite{sheppard}.
%However, the dependence of the internal beating
%  frequency on the soliton parameters is unclear.
However, and since the present study focused predominantly
on the five-parameter family stemming from the SO(2) rotation, the
  persistence, stability, and interactions of the full
seven parameter solitonic states (and even the eight parameter
generalization thereof presented within~\cite{parkshin})
  in the non-integrable case constitute themes worthy of further study.

There are
%numerous
many other directions that are worth considering further along the
lines of this work.
%this track.
Quantifying further (and semi-analytically, if possible)
the interactions between the beating dark-dark
solitons, as well as studying in more detail the dark-antidark solitons
%entities
that appear to spontaneously arise from their breakup in the miscible
regime are interesting
%some among the one-dimensional
extensions of this work in the one-dimensional setting.
%that appear to be of interest.
On the other hand, one naturally may consider
the two-dimensional (2D) generalization of the considerations herein,
especially upon bearing in mind that the SU(2) (or SO(2)) rotations
used herein are not restricted to the one-dimensional realm in any
particular way. In that regard, one may envision vortex-bright
soliton states~\cite{kody} (i.e., the 2D analog of the dark-bright
waves) rotated via SO(2) to produce vortex-vortex type states
(in analogy to the dark-dark ones). Such states are currently
under study and will be reported in future publications.

\acknowledgments P.G.K.\ acknowledges the support from NSF-DMS-0806762 and from the
Alexander von Humboldt Foundation.
The work of D.J.F. was partially supported by the Special Account for
Research Grants of the University of Athens.
P.E. acknowledges
support from NSF and ARO.
M.H. acknowledges support from NSF DMS 1008973.
J.C. acknowledges financial support from the MICINN project FIS2008-04848.


\begin{thebibliography}{99}
\bibitem{note} Note that we cannot do this analysis for systems with
  $g_{11}:g_{12}:g_{22} \ne 1$
%a ratio between scattering lengths different to one,
as, in that case, there is always an oscillation of the center of mass
(alike to that of a particle in a well) with a frequency non-commensurable to the beating frequency.
Consequently,
%there are no
periodic orbits do not exist.
\end{thebibliography}
\end{document}